\documentclass[11pt,a4paper]{article}
\usepackage{cite}
\usepackage{graphicx}
\usepackage{amssymb}
\usepackage{amsmath}
\usepackage{amsfonts}
\usepackage{dsfont}
\usepackage{mathtools}
\usepackage{rotating}
\usepackage{bbold,amsfonts}

\usepackage[utf8]{inputenc}
\usepackage{bm}
\usepackage{xcolor}
\usepackage{braket}

\usepackage[height=8.8in,width=6.45in]{geometry}
\usepackage[font=small,labelfont=bf]{caption}
\usepackage[hidelinks]{hyperref}
\usepackage{booktabs,float,slashed}
\numberwithin{equation}{section}
\newcommand{\vev}[1]{{\left\langle #1 \right\rangle}}

\newcommand{\beq}{\begin{equation}}
\newcommand{\eeq}{\end{equation}}

\newcommand{\overbar}[1]{\mkern 1.5mu\overline{\mkern-1.5mu#1\mkern-1.5mu}\mkern 1.5mu}

\newcommand{\p}{\pi}

\DeclareMathOperator{\tr}{tr}

\newcommand{\ii}{\mathrm{i}}

\makeatletter
\newcommand*{\letterdef@}{}
\newcommand*{\letterdef}[3]{%
	\def\letterdef@##1{\expandafter\newcommand\csname #1\endcsname{#2{##1}}}%
	\@tfor\@tempa :=#3\do{\expandafter\letterdef@\expandafter{\@tempa}}}
\makeatother
\letterdef{c#1} {\mathcal}{ABCDEFGHIJKLMNOPQRSTUVWXYZ} 
\letterdef{rm#1}{\mathrm} {dDeimM} 

\newcommand{\Xx}{\mathsf{X}}
\newcommand{\Dx}{\mathsf{D}}
\newcommand{\Sx}{\mathsf{S}}
\newcommand{\Yx}{\mathsf{Y}}
\newcommand{\Nx}{\mathsf{N}}
\newcommand{\Lax}{\mathsf{\Lambda}}

\newdimen\tableauside\tableauside=1.0ex
\newdimen\tableaurule\tableaurule=0.4pt
\newdimen\tableaustep
\def\phantomhrule#1{\hbox{\vbox to0pt{\hrule height\tableaurule
			width#1\vss}}}
\def\phantomvrule#1{\vbox{\hbox to0pt{\vrule width\tableaurule
			height#1\hss}}}
\def\sqr{\vbox{%
		\phantomhrule\tableaustep
		\hbox{\phantomvrule\tableaustep\kern\tableaustep\phantomvrule\tableaustep}%
		\hbox{\vbox{\phantomhrule\tableauside}\kern-\tableaurule}}}
\def\squares#1{\hbox{\count0=#1\noindent\loop\sqr
		\advance\count0 by-1 \ifnum\count0>0\repeat}}
\def\tableau#1{\vcenter{\offinterlineskip
		\tableaustep=\tableauside\advance\tableaustep by-\tableaurule
		\kern\normallineskip\hbox
		{\kern\normallineskip\vbox
			{\gettableau#1 0 }%
			\kern\normallineskip\kern\tableaurule}%
		\kern\normallineskip\kern\tableaurule}}
\def\gettableau#1 {\ifnum#1=0\let\next=\null\else
	\squares{#1}\let\next=\gettableau\fi\next}
\allowdisplaybreaks

\begin{document}
\begin{titlepage}
\vspace*{10mm}
\begin{center}
{\LARGE \bf 
	 	Exact results in a $\cN = 2$ superconformal gauge theory at strong coupling
}

\vspace*{15mm}

{\Large M. Beccaria${}^{\,a}$, M. Bill\`o${}^{\,b,d}$, M. Frau${}^{\,b,d}$, A. Lerda${}^{\,c,d}$, A. Pini${}^{\,d}$}

\vspace*{8mm}
	
${}^a$ Università del Salento, Dipartimento di Matematica e Fisica ``Ennio De Giorgi'',\\ 
		and I.N.F.N. - sezione di Lecce, \\Via Arnesano, I-73100 Lecce, Italy
			\vskip 0.3cm
			
${}^b$ Universit\`a di Torino, Dipartimento di Fisica,\\
			Via P. Giuria 1, I-10125 Torino, Italy
			\vskip 0.3cm
			
${}^c$  Universit\`a del Piemonte Orientale,\\
			Dipartimento di Scienze e Innovazione Tecnologica\\
			Viale T. Michel 11, I-15121 Alessandria, Italy
			\vskip 0.3cm
			
${}^d$   I.N.F.N. - sezione di Torino,\\
			Via P. Giuria 1, I-10125 Torino, Italy

\vskip 0.8cm
	{\small
		E-mail:
		\texttt{matteo.beccaria@le.infn.it, billo,frau,lerda,pini@to.infn.it}
	}
\vspace*{0.8cm}
\end{center}

\begin{abstract}
We consider the $\cN=2$ SYM theory with gauge group SU($N$) and a matter content 
consisting of one multiplet in the symmetric and one in the anti-symmetric representation. 
This conformal theory admits a large-$N$ 't Hooft expansion and is dual to a particular orientifold 
of $\mathrm{AdS}_{5}\times S^{5}$.
We analyze this gauge theory relying on the matrix model provided by localization \`a la Pestun. 
Even though this matrix model has very non-trivial interactions, 
by exploiting the full Lie algebra approach to the matrix integration, we show that
a large class of observables can be expressed in a closed form in terms of an infinite matrix
depending on the 't Hooft coupling $\lambda$. 
These exact expressions can be used to generate the perturbative expansions at high orders in a very efficient way, and also to study analytically the leading behavior at strong coupling. 
We successfully compare these predictions to a direct Monte Carlo numerical evaluation of the matrix integral and to the Pad\'e resummations derived from very long perturbative series, 
that turn out to be extremely stable beyond the convergence disk $|\lambda|<\pi^2$ of the latter.
\end{abstract}
\vskip 0.5cm
	{
		Keywords: {$\mathcal{N}=2$ conformal SYM theories, strong coupling, matrix model}
	}
\end{titlepage}
\setcounter{tocdepth}{2}
\tableofcontents
\vspace{1cm}

\section{Introduction and summary of results}
\label{secn:intro}
Four-dimensional gauge theories with extended supersymmetry are intensively explored
to obtain exact results in quantum field theory and improve our understanding of the strong-coupling regime. Important progress in these directions has been achieved, over the years, in the case of $\cN=4$ SYM theories by exploiting integrability, localization techniques and the AdS/CFT correspondence.

The analysis of cases with reduced supersymmetry is of course more difficult, but for $\cN=2$ theories significant steps have been taken.
In particular, for these theories localization techniques can still be used and some instances of holographic correspondence are known. In particular, as shown in \cite{Pestun:2007rz},
by putting a $\cN=2$ SYM theory on a compact space like a four-sphere, one can localize 
the infinite-dimensional path-integral on a finite-dimensional locus and reduce the calculation 
to an interacting matrix model. If conformal symmetry is present\,%
\footnote{$\cN=2$ superconformal theories were originally investigated in \cite{Howe:1983wj}.}, 
this matrix model also encodes information on the observables of
the theory in flat space. In this way the partition function and the vacuum expectation value of 
the BPS Wilson loop have been computed \cite{Pestun:2007rz,Andree:2010na,Rey:2010ry,Passerini:2011fe,Bourgine:2011ie,Russo:2012ay,Beccaria:2021vuc}.
Many other observables of the gauge theory can be described by the matrix model, such as the extremal correlators of chiral/anti-chiral fields \cite{Baggio:2014ioa,Baggio:2014sna,Gerchkovitz:2014gta,Baggio:2015vxa,Fiol:2015mrp,Gerchkovitz:2016gxx,Rodriguez-Gomez:2016ijh,Rodriguez-Gomez:2016cem,Baggio:2016skg,Billo:2017glv,Bourget:2018obm,Bourget:2018fhe,Billo:2019job,Beccaria:2020hgy}, the correlators of chiral operators with the BPS Wilson loop \cite{Rodriguez-Gomez:2016cem,Billo:2018oog,Beccaria:2018owt,Billo:2019fbi} and the brehmsstrahlung function \cite{Fiol:2015spa,Mitev:2015oty,Gomez:2018usu,Bianchi:2019dlw}. 
The same localization techniques can also be applied to study various properties of the massive 
$\cN=2^*$ SYM theory \cite{Buchel:2013id,Russo:2013qaa,Russo:2013kea,Billo:2014bja} and of quiver theories \cite{Pini:2017ouj,Zarembo:2020tpf,Galvagno:2020cgq,Beccaria:2021ksw,Galvagno:2021bbj}.

These matrix-model computations can be carried out following two different strategies, called the 
Cartan sub-algebra and the full Lie algebra approach in \cite{Fiol:2020bhf}. In the first approach, the matrix model is written as an integral over the matrix eigenvalues and its large-$N$ limit is described by the asymptotic eigenvalue distribution determined within a saddle-point approximation.
The second approach, instead, keeps the matrix integral over the full Lie algebra and exploits recursion relations \cite{Billo:2017glv} to evaluate correlation functions. This technique, originally developed because of its effectiveness at finite $N$, has been applied also to study the 
large-$N$ and the
large-charge sectors of gauge theories \cite{Beccaria:2018xxl,Fiol:2018yuc,Grassi:2019txd,Beccaria:2020azj}.

For the $\cN=4$ SYM theory with group SU($N$) the matrix model arising from localization is Gaussian, which makes relatively easy to treat it both at finite $N$ and in the large-$N$ limit, and also to discuss its strong-coupling behavior. On the contrary, for $\cN=2$ SYM theories 
the matrix model has a very complicated potential and its use for an analytical description of the strong-coupling regime is far from obvious. 

Previous strong-coupling results have been obtained in the Cartan sub-algebra approach. In particular, in the seminal work 
\cite{Passerini:2011fe} the expectation value of the BPS Wilson loop in $\cN=2$ SYM with $N_F=2N$ fundamental flavors (SQCD) has been computed in the planar approximation
at leading order for large values of the 't Hooft coupling $\lambda$, by solving 
an associated Wiener-Hopf problem. 
The planar free energy of SQCD has been later evaluated in \cite{Russo:2012ay} and generalizations
to $\cN=2$ superconformal theories with gauge group SO($N$) or Sp($N$) have been obtained in 
\cite{Fiol:2015mrp}. An important outcome
is that the planar expectation value of the Wilson loop scales proportionally to $\lambda^{\eta}$ with an exponent $\eta$ determined by the  large-$N$ limit of the ratio $N_{F}/N$, confirming planar equivalence
with $\cN=4$ SYM (where $N_{F}=0$) in all models where $N_{F}$ does not grow with $N$.
The strong-coupling limit of the chiral two-point functions at planar level has been initially
studied in \cite{Rodriguez-Gomez:2016ijh} and further explored in \cite{Baggio:2016skg}. More recently, the Wiener-Hopf method has been successfully exploited to determine the strong-coupling expectation value of the BPS Wilson loop in the SU($N$)$\times$SU($N$) quiver theory
\cite{Zarembo:2020tpf}. Although still at planar level, this calculation is at the 
next-to-leading order in the 't Hooft coupling, and represents an important extension of the 
leading-order results of \cite{Rey:2010ry,Passerini:2011fe}. Despite attempts to simplify the mathematical structure of the Wiener-Hopf approach (see for example \cite{Bourgine:2011ie}), this method has not been extended beyond the planar level, the main technical obstruction being that it is not based on a controlled expansion in terms of a parametrically small quantity.

The study of the strong-coupling regime in the large-$N$ limit for $\cN=2$ superconformal 
theories is clearly an important ingredient also to achieve a full understanding of the AdS/CFT correspondence in systems with a reduced amount of supersymmetry, at least for those models
that possess an holographic dual that can be accessed at strong coupling from the gauge side.

To address these issues it is convenient to consider $\cN=2$ theories that are as close as possible to $\cN=4$ SYM. Therefore, in this paper we focus on a particular 
$\cN=2$ conformal model with gauge
group SU($N$), called $\mathbf{E}$ theory in \cite{Billo:2019fbi,Beccaria:2020hgy}, which 
is very close to $\cN=4$ SYM in many ways. In the 't Hooft limit, it possesses a large class of observables, comprising the free energy and the expectation value of the BPS Wilson loop, which agree with the corresponding ones in the $\cN=4$ SYM at the planar level
\cite{Fiol:2020bhf,Beccaria:2020hgy} and deviate from the $\cN=4$ results 
only at order $1/N^2$. Moreover, it admits a holographic dual of the form
$\mathrm{AdS}_5\times S^5/\mathbb{Z}_2$ which is obtained by means of a $\mathbb{Z}_2$ 
orbifold/orientifold procedure from the holographic dual of $\cN=4$ SYM (see for instance \cite{Ennes:2000fu}). Actually, the $\mathbf{E}$ theory itself can be realized by taking a suitable 
orientifold projection of a two-node quiver model, which in turn can be engineered with a system of fractional D3-branes in a $\mathbb{Z}_2$ orbifold \cite{Park:1998zh}.
It is natural to distinguish the observables of the $\mathbf{E}$ theory in two classes, those corresponding to operators that belong to the untwisted sectors in the string construction, and those associated to operators that belong to the twisted sectors. For this reason, throughout this paper we will use the terminology of ``untwisted'' and ``twisted'' observables.

In the holographic dictionary an untwisted observable of the $\mathbf{E}$ theory, which we generically denote by $\big\langle \,\cU\,\big\rangle$, is described by a supergravity excitation of $\mathrm{AdS}_5\times S^5$ which is even under the orbifold/orientifold parity. Therefore, the same excitation exists also in the maximally supersymmetric case where it corresponds to an observable of the $\cN=4$ SYM theory which we may call $\big\langle\, \cU\,\big\rangle_0$. One then expects that the 
$\mathbf{E}$-theory result be identical to the one in $\cN=4$ at the leading order in the large-$N$ expansion, with a deviation starting at the non-planar level:
\begin{equation}
\frac{\big\langle \,\cU\,\rangle}{\big\langle\, \cU\,\rangle_0}=
1+\frac{\delta\,\cU}{N^2}+O(N^{-4})~,
\label{untwisted}
\end{equation}
where $\delta\,\cU$ is a non-trivial function of the 't Hooft coupling. The free energy and the vacuum expectation value of the BPS Wilson loop are examples of untwisted observables \cite{Fiol:2020bhf,Beccaria:2020hgy,Beccaria:2021vuc} but, as we will see, also the correlators of gauge invariant single-trace operators with even dimension in the $\mathbf{E}$ theory
belong to the untwisted class.

In the $\cN=4$ theory there are observables, which we call $\big\langle\, \cT\,\big\rangle_0$,
that in the holographic correspondence are mapped to supergravity excitations of 
$\mathrm{AdS}_5\times S^5$ which are odd under the $\mathbb{Z}_2$ orbifold/orientifold parity. These excitations
are removed by the projection and get replaced by states of the twisted sectors corresponding
to the twisted observables $\big\langle\, \cT\,\big\rangle$ of the $\mathbf{E}$ theory. 
Therefore, one expects that $\big\langle\, \cT\,\big\rangle$ and $\big\langle\, \cT\,\big\rangle_0$
differ already at the planar level:
\begin{equation}
\frac{\big\langle \,\cT\,\rangle}{\big\langle \,\cT\,\rangle_0}=
1+\delta\,\cT+O(N^{-2})~,
\label{twisted}
\end{equation}
where $\delta\,\cT$ depends non-trivially on the 't Hooft coupling $\lambda$. As already discussed
in \cite{Beccaria:2020hgy}, the correlators involving of gauge invariant single-trace operators 
with odd dimension in the $\mathbf{E}$ are of twisted type and behave as in (\ref{twisted}).

The main goal of this paper is to provide exact results for the functions $\delta\,\cU$ and
$\delta\,\cT$ for a large class of observables, to exhibit their exact dependence on the 
't Hooft coupling $\lambda$, and to find eventually their strong-coupling behavior when $\lambda\to \infty$.

\subsection{Results}
By exploiting the simplicity of the matrix model and relying on the power of the full Lie algebra approach, we have analyzed in detail several observables of the $\mathbf{E}$ theory in the
large-$N$ limit in the untwisted and in the twisted sectors, and found the leading corrections with respect to the $\cN=4$ theory using a simplified set of recursion relations \cite{Beccaria:2020hgy}.
As a consequence, we show that these quantities can be effectively
evaluated in a Gaussian matrix model whose quadratic form is an infinite $\lambda$-dependent
matrix
\begin{equation}
\mathbb{1}-\Xx~,
\label{matrix0}
\end{equation}
whose elements are convolutions of Bessel functions of the first kind. More precisely, we find that
for the untwisted observables we have considered, 
the deviation $\delta\,\cU$ is given in terms of logarithmic 
$\lambda$-derivatives of
\begin{equation}
\log\det\big(\mathbb{1}-\Xx\big)~,
\end{equation}
which is proportional to the free energy of the effective Gaussian model, while for
the twisted variables the deviation $\delta\,\cT$ is expressed in closed form in terms of the propagator
\begin{equation}
\big(\mathbb{1}-\Xx\big)^{-1}~.
\end{equation}
Since the matrix $\Xx$ is exactly known as a function of $\lambda$ through the Bessel functions entering its definition, we can expand these expressions for small values of $\lambda$ and
obtain the weak-coupling expansions. Actually, proceeding in this way, we are able to push the
perturbative calculations to very high orders with relatively low computational effort and generate
very long expansions that can then be used for numerical investigations. 

Most importantly, we can exploit the exact knowledge of the $\lambda$-dependence of the $\Xx$ matrix and study the above deviations at strong coupling in an analytic way. For  
$\lambda\to\infty$ we find that they behave as
\begin{equation}
\begin{cases}
\,\sqrt{\lambda}\quad\mbox{\,for~the~untwisted~observables}~,\\[1mm]
\,{\lambda^{-1}}\quad\mbox{for~the~twisted~observables}~.
\end{cases}
\end{equation}
The precise numerical coefficients in front of $\lambda^{-1}$ for the twisted observables we have analyzed can be fixed in full generality, while the coefficients in front of $\sqrt{\lambda}$ for the
untwisted observables are more delicate and depend on the precise strong-coupling behavior of the free energy. However, once this is fixed, also these coefficients are fixed.

To substantiate our findings, we have compared the analytical strong-coupling predictions with numerical results obtained by using Pad\'e approximants of the long perturbative series and by an independent Monte Carlo simulation. The agreement we have found is remarkable, especially in the twisted case.

Since the $\mathbf{E}$ theory has a holographic dual corresponding to a geometry of the type
$\mathrm{AdS}_5\times S^5/\mathbb{Z}_2$, it would very interesting to retrieve the above results
from a gravitational point of view. In particular, our results for the twisted
observables provide a precise prediction for the twisted sectors of the $\mathrm{AdS}_5\times S^5/\mathbb{Z}_2$ geometry, which so far have remained largely unexplored. 
Furthermore, it would be interesting to go beyond the leading order in the strong-coupling expansion and study the properties of the resulting series for large $\lambda$, 
and also to investigate the role played by non-perturbative instanton corrections and/or resurgence effects, along the lines recently discussed in \cite{Dorigoni:2021bvj,Dorigoni:2021guq} 
for the $\cN=4$ SYM theory.

\subsection{Outline of the paper}
In Section~\ref{secn:matrix} we review the main properties of the $\mathbf{E}$ theory and of its
associated matrix model arising from localization. In Section~\ref{sec:leading} we derive the exact expressions of the leading corrections in the large-$N$ expansion for twisted and untwisted
observables of the $\mathbf{E}$ theory in terms of the infinite matrix $\Xx$. In 
Section~\ref{sec:large} we work out the analytic behavior at strong coupling of these observables
at leading order, which we check against numerical analyses and Monte Carlo simulations in 
Section~\ref{sec:numerical}. Finally, a few more technical details concerning the infinite matrix $\Xx$ and its properties are collected in Appendix~\ref{app:details}.

\section{$\mathcal{N}=2$ conformal SYM theories and matrix model}
\label{secn:matrix}

In this section we review some key properties of the $\cN=2$ conformal theories with the aim of establishing the notation and describing the set-up 
in which we will perform our calculations. Most of this material has already been discussed in the 
literature (see in particular \cite{Billo:2019fbi,Beccaria:2020hgy}) and thus we can be 
rather brief.

\subsection{$\mathcal{N}=2$ conformal SYM theories}
We consider $\mathcal{N}=2$ SYM theories in 4d with gauge group SU($N$) whose 
matter content consists of $N_F$ hypermultiplets in the fundamental representation, 
$N_S$ hypermultiplets in the symmetric representation and $N_A$ hypermultiplets in the
anti-symmetric representation. These theories are conformal provided the 1-loop
$\beta$-function coefficient 
\begin{equation}
\beta_0=2N-N_F-N_S(N+2)-N_A(N-2)
\label{beta0}
\end{equation}
vanishes. Among the five families of models that satisfy this condition 
(see \cite{Howe:1983wj,Koh:1983ir,Bhardwaj:2013qia} and also 
\cite{Fiol:2015mrp,Bourget:2018fhe,Billo:2019fbi,Beccaria:2020hgy}), 
here we focus on the following two families, parametrized by $N$ and 
characterized by
\begin{equation}
\begin{aligned}
\mathbf{D}~&:~~N_F=4~,~~~N_S=0~,~~~N_A=2~,\\[1mm]
\mathbf{E}~&:~~N_F=0~,~~~N_S=1~,~~~N_A=1~.
\end{aligned}
\label{DE}
\end{equation}
Even though these theories have received less attention than SQCD, which has $N_F=2N$, 
$N_S=N_A=0$, 
they are very interesting for various reasons. Most notably, they share 
many features with $\mathcal{N}=4$ SYM in the 
large-$N$ limit, and admit a holographic dual description in terms of strings 
propagating in a geometry of the form $\mathrm{AdS}_5\times S^5/\mathbb{Z}_2$, 
where $\mathbb{Z}_2$ is a suitable orientifold group (see for example 
\cite{Ennes:2000fu}).

For both $\mathbf{D}$ and $\mathbf{E}$ families, the ratio of the number of 
fundamental flavors with respect to (twice) 
the number of colors scales to zero in the planar limit, namely
\begin{equation}
\nu=\lim_{N\to\infty}\frac{N_F}{2N}=0~,
\label{nu}
\end{equation}
while in SQCD one has $\nu=1$. The vanishing of $\nu$ is a necessary condition for planar equivalence to $\cN=4$ SYM,  at least in a certain ``even'' sector \cite{Fiol:2020bhf,Beccaria:2020hgy}.

Another interesting feature is that the difference between the $c$ and $a$ 
central charges is such that
\begin{equation}
\begin{aligned}
\alpha_0&=\lim_{N\to\infty}\frac{c-a}{N^2}=0~~\mbox{for}~\mathbf{D}~
\mbox{and}~\mathbf{E}~,\\[2mm]
\alpha_1&=\lim_{N\to\infty}\frac{c-a}{N}=\begin{cases}
\frac{1}{8}~&\mbox{for}~\mathbf{D}~,\\[1mm]
0~&\mbox{for}~\mathbf{E}~,
\end{cases}
\\[2mm]
\alpha_2&=\lim_{N\to\infty}(c-a)=\begin{cases}
\infty~&\mbox{for}~\mathbf{D}~,\\[1mm]
\frac{1}{24}~&\mbox{for}~\mathbf{E}~.
\end{cases}
\end{aligned}
\label{ca}
\end{equation}
Again this is quite different from SQCD, for which $\alpha_0=\frac{1}{24}$ and
$\alpha_1=\alpha_2=\infty$.

On the other hand, the $\mathcal{N}=4$ SU($N$) SYM theory  has $\nu=0$, since there 
is no fundamental matter, and $\alpha_0=\alpha_1=\alpha_2=0$, since the two central
charges are equal, $c=a$. We therefore see that in the large-$N$ limit the 
$\mathbf{D}$ and $\mathbf{E}$ families share many features with the 
$\mathcal{N}=4$ SYM theory, and in particular the $\mathbf{E}$ theory is the one
which goes closer. 
In this sense the $\mathbf{E}$ theory can be considered as 
the next-to-simplest SYM theory after the maximally supersymmetric one. 
From now on we will discuss only the $\mathbf{E}$ theory, even if most of 
our considerations can be applied with minor modifications to the $\mathbf{D}$ theory 
as well.

In the $\mathcal{N}=2$ SYM theories, an important set of local operators is provided by 
the multitraces  
\begin{align}
	\label{defOn}
		O_{\mathbf{n}}(x) \equiv \tr \varphi^{n_1}(x)\, \tr \varphi^{n_2}(x) \ldots
\end{align}
where $\mathbf{n}=\{n_i\}$ and $\varphi$ is the (complex) scalar field in the adjoint 
vector multiplet.
These operators are gauge-invariant, chiral, possess an $R$-charge equal to $n = \sum_i n_i$ and are 
automatically normal-ordered because of $R$-charge conservation. 
The anti-chiral operators $\overbar O_{\mathbf{n}}(x)$ are constructed in the same way 
with the conjugate field $\overbar \varphi(x)$. In the conformal case, the two-point 
functions between chiral and anti-chiral operators take the general form
\begin{equation}
	\label{twopointdef}
		\big\langle O_{\mathbf{n}}(x) \,\overbar O_{\mathbf{m}}(0)\big\rangle 
		= \frac{G_{\mathbf{n},\mathbf{m}}}{(4\pi^2 x^2)^{m+n}\phantom{\big|}}\,
		\delta_{m,n}
\end{equation}
where $G_{\mathbf{n},\mathbf{m}}$ is a non-trivial function of the gauge coupling $g$ 
and of the number of colors $N$.
To keep the presentation as simple as possible, in the following we restrict our attention to the single trace operators
\begin{align}
	\label{defOns}
		O_{n}(x) \equiv \tr \varphi^{n}(x)
\end{align}
in which case the coefficients in the two-point functions (\ref{twopointdef}) become 
diagonal and are just denoted by $G_n$. These coefficients can be studied in the 
limit $N\to\infty$ with the 't Hooft coupling $\lambda=g^2N$ held fixed, and compared 
with the corresponding coefficients in the $\mathcal{N}=4$ SYM theory, denoted by 
$G_n^{(0)}$. For the $\mathbf{E}$ theory one finds
\begin{subequations}
\begin{align}
\frac{G_{2k+1}\phantom{\big|}}{G_{2k+1}^{(0)}\phantom{\big|}}&=1+\Delta_k(\lambda)+\mathcal{O}(N^{-2})~,
\label{Deltaodd}
\\[1mm]
\frac{G_{2k}\phantom{\big|}}{G_{2k}^{(0)}\phantom{\big|}}&=
1+\frac{\delta_k(\lambda)}{N^2}+\mathcal{O}(N^{-4})~,
\label{Deltaeven}
\end{align}
\label{Deltas}%
\end{subequations}
where $\Delta_k$ and $\delta_k$ are non-trivial functions of the 't Hooft coupling only. 
This means that when $N\to\infty$ the two-point functions of even 
operators are identical to those in the $\mathcal{N}=4$ theory and differ only in
sub-leading terms at order $1/N^2$, while the two-point functions of odd 
operators are different from the $\mathcal{N}=4$ 
ones already in the planar limit\,%
\footnote{A similar result is found in the $\mathbf{D}$ theory,
but the first corrections that start at order $1/N$ instead of $1/N^2$.}.

This structure is in full agreement with the expectations based on the AdS/CFT 
correspondence. Indeed, as discussed in the Introduction (see (\ref{untwisted}) and (\ref{twisted})), 
when the holographic dictionary is applied to the $\mathrm{AdS}_5\times S^5/\mathbb{Z}_2$ orientifold that is dual to the 
$\mathbf{E}$ theory, the even chiral operators $O_{2k}$ map to 
supergravity states of $\mathrm{AdS}_5\times S^5$ which survive the 
projection. In other words, these excitations belong to the untwisted sectors of the 
orientifold and thus in the planar limit their correlators must be the same as in the
maximally supersymmetric case. On the contrary, the $\mathrm{AdS}_5\times S^5$ 
states that correspond to the odd chiral operators in the $\cN=4$ theory are projected 
out in the orientifold construction, and the odd chiral operators $O_{2k+1}$ of the 
$\cN=2$ theory are obtained from states of the twisted sectors of 
$\mathrm{AdS}_5\times S^5/\mathbb{Z}_2$. These supergravity excitations have no 
analogue in $\mathrm{AdS}_5\times S^5$, and thus it has to be expected that even in 
the planar limit their correlators be different from those in the $\cN=4$ theory.

Our aim is to find an explicit expression of the functions $\Delta_k(\lambda)$ 
and $\delta_k(\lambda)$ 
appearing in (\ref{Deltas}). One 
possibility to reach this goal is to use the standard perturbative field-theory methods and
compute the Feynman diagrams that contribute to the two-point functions.
This method is conceptually straightforward, but in practice it quickly becomes rather
daunting as the number of loops increases. Even if some simplifications may occur
in the large-$N$ limit, still the number of diagrams that one has to compute is 
tremendously high (see \cite{Billo:2019fbi,Beccaria:2020hgy} for some explicit
examples at the lowest perturbative orders). Thus only the very first few perturbative
terms can be explicitly computed in this way.

A much better approach is to exploit localization and map the superconformal gauge theory
to an equivalent matrix model \cite{Pestun:2007rz}.

\subsection{Matrix model}

If one defines the $\cN=2$ SYM theory on a sphere $S^4$ of unit radius, 
its partition function can be expressed in terms of a matrix model as follows 
\cite{Pestun:2007rz}\,%
\footnote{As compared to \cite{Pestun:2007rz}, we have rescaled the matrix $a$ with a factor of
$\sqrt{\frac{g^2}{8\pi^2}}$ in order to obtain a canonical normalization in the quadratic term, and
have neglected the overall factor $\Big(\frac{g^2}{8\pi^2}\Big)^{\frac{(N^2-1)}{2}}$ since it drops out in all calculations.}:
\begin{equation}
	\label{intda}
		\cZ = \int \!da~\rme^{-\tr a^2}\,|Z_{\mathrm{1-loop}}\,Z_{\mathrm{inst}}|^2~. 
\end{equation}
Here we have adopted the so-called ``full Lie algebra approach" \cite{Billo:2017glv,Billo:2019fbi,Beccaria:2020hgy} in which the integration is performed
over all components of a Hermitean traceless matrix $a$ belonging to the Lie algebra\,%
\footnote{We write $a=a^b \,t_b$ ($b=1,\ldots,N^2-1$) where $t_b$ are the generators of SU($N$)
normalized in such a way that $\tr t_b\, t_c = \frac 12\, \delta_{bc}$} of SU($N$). 
The factor $Z_{\mathrm{1-loop}}$ arises from a 1-loop computation, while
$Z_{\mathrm{inst}}$ is the non-perturbative instanton contribution. In the large-$N$ 't Hooft limit, instantons are suppressed and thus we can put $Z_{\mathrm{inst}}=1$.

The 1-loop term depends on the representation $\cR$ of the matter multiplets in the gauge theory
and takes the following form
\begin{equation}
|Z_{\mathrm{1-loop}}|^2=\rme^{-S_{\mathrm{int}}(a)}
\end{equation}
with
\begin{equation}
S_{\mathrm{int}}(a)=\tr_{\cR} \log H\Big(\ii\sqrt{\frac{g^2}{8\p^2}} a\Big) - 
\tr_{\mathrm{adj}} \log H\Big(\ii\sqrt{\frac{g^2}{8\p^2}} a\Big)~.
\label{StologH}
\end{equation}
Here
\begin{equation}
H(x)=G(1+x)\,G(1-x)~,
\label{His}
\end{equation}
with $G$ being the Barnes $G$-function. In the $\cN=4$ SYM theory, where $\cR$ is the adjoint representation, $S_{\mathrm{int}}$ clearly vanishes, while in the $\cN=2$ SYM theories it accounts 
for the matter content of the so-called ``difference theory'' \cite{Andree:2010na}, 
in which one removes from the $\cN=4$ result the contributions of the adjoint hypermultiplets 
and replaces them with the contributions of hypermultiplets in the representation $\cR$.
In the perturbative regime when $g\to 0$ we can use the expansion
\begin{equation}
\log H(x)=- (1+\gamma_{\mathrm{E}}) x^2-\sum_{p=1}^\infty\frac{\zeta(2p+1)}{p+1}\,x^{2p+2}~,
\label{expH}
\end{equation}
where $\gamma_{\mathrm{E}}$ is the Euler-Mascheroni constant,
and check that for the $\mathbf{E}$ theory $S_{\mathrm{int}}$ is given by
\cite{Billo:2019fbi,Beccaria:2020hgy}
\begin{equation}
S_{\text{int}}(a)= 4\sum_{\ell,m=1}^\infty (-1)^{\ell+m}
\Big(\frac{g^2}{8\pi^2}\Big)^{\ell+m+1}
\frac{(2\ell+2m+1)!}{(2\ell+1)!\,(2m+1)!}\,\zeta(2\ell+2m+1)\,
\tr a^{2\ell+1}\,\tr a^{2m+1}~.
\label{Sint}
\end{equation}
Therefore, the partition function (\ref{intda}) can be written as
\begin{equation}
\cZ = \int \!da~\rme^{-\tr a^2}\,\rme^{-S_{\mathrm{int}}(a)}~,
\label{ZS4}
\end{equation} 
and $S_{\mathrm{int}}(a)$ can be interpreted as an interaction action. Note that even for the $\mathbf{E}$ theory, which is the simplest $\cN=2$ conformal gauge theory as we mentioned before, the corresponding matrix model contains 
an infinite number of interactions as is clear from (\ref{Sint}). This is to be contrasted with the matrix model 
corresponding to the $\cN=4$ SYM theory which is purely Gaussian. 

Given any function $f(a)$, its vacuum expectation value in the $\cN=2$ matrix model is
\begin{align}
	\label{vevmat}
		\big\langle f(a) \big\rangle \,
		= \,\frac{\displaystyle{ \int \!da ~f(a)\,\rme^{-\tr a^2-S_\mathrm{int}(a)}}}
		{ \displaystyle{\int \!da~\rme^{-\tr a^2-S_\mathrm{int}(a)}}}	
		= \frac{\displaystyle{\big\langle\,
		f(a)\,\rme^{- S_\mathrm{int}(a)}\,\big\rangle_{0}\phantom{\Big|}}}{\displaystyle{\big\langle\,
		\rme^{- S_\mathrm{int}(a)}\,\big\rangle_{0}\phantom{\Big|}}}~,
\end{align} 
where in the second step we have denoted by $\langle~~\rangle_{0}$ the expectation value 
in the free Gaussian model.
In particular, we are interested in the matrix operators $O_n(a)$ that correspond to the chiral operators (\ref{defOns}) of the gauge theory and in their correlators. Indeed,
as shown in \cite{Baggio:2014ioa,Baggio:2014sna,Gerchkovitz:2014gta,Baggio:2015vxa,Baggio:2016skg,Gerchkovitz:2016gxx,Rodriguez-Gomez:2016ijh,Rodriguez-Gomez:2016cem,Billo:2017glv}, the coefficients $G_n$ in the two-point functions
of the gauge theory in flat space are given by
\begin{equation}
G_n=\big\langle O_n(a)\,O_n(a)\big\rangle~.
\label{Gnmat}
\end{equation}
In this approach everything is then reduced to the calculation of vacuum expectation values in the free matrix model using (\ref{vevmat}). As discussed in \cite{Billo:2017glv,Billo:2019fbi,Beccaria:2020hgy} these vacuum expectation values can be efficiently computed using recursion relations, and explicit expressions can be easily generated even at very high perturbative orders.

\section{Leading corrections in the 't Hooft limit}
\label{sec:leading}
Our aim is to provide exact formal expressions of the leading corrections in the large-$N$ limit
for both twisted and untwisted operators in the $\mathbf{E}$ matrix model and for the 
corresponding observables in the gauge theory. 
In this section we show that these formal expressions can be encoded in an auxiliary theory of 
infinite free variables. In particular, we prove that the twisted corrections involve the propagator of this free theory, while the untwisted ones are expressed as derivatives 
with respect to $\lambda$ of its free energy. 
From such exact expressions it is quite straightforward to expand in powers of $\lambda$ and obtain the perturbative results at very high orders. From these long perturbative series it is then possible to compute Pad\'e approximants which remain stable and accurate well beyond the radius of convergence of the perturbative series, thus allowing for an extrapolation towards strong coupling.
In the next section we will describe how to extract the leading behavior of the corrections at large 
$\lambda$ in an analytic way. To test the correctness of the results, in Section \ref{sec:numerical} we will make a comparison with the direct numerical evaluation of the vacuum expectation values in the matrix model by means of Monte Carlo simulations extrapolated to the large-$N$ limit.   

\subsection{Gaussian representation of the $\mathbf{E}$ model at large $N$}
\label{subsec:ff}
Let us first recapitulate a few results of \cite{Beccaria:2020hgy}. Consider the expectation 
value of a product of an even number of odd traces in the Gaussian theory:
\begin{align}
\label{toddis}
		\big\langle\tr a^{2k_1+1}\, \tr a^{2k_2+1}\, \tr a^{2k_3+1}\, \tr a^{2k_4+1}  \cdots\big\rangle_0 \,\equiv\, 
t_{2k_1+1,2k_2+1,2k_3+1,2k_4+1,\cdots} ~.
\end{align} 
At the leading large-$N$ order we have
\begin{align}
	\label{tracewick}
	t_{2k_1+1,2k_2+1,2k_3+1,2k_4+1,\cdots} 
		& = \prod_i\Big(\frac{N^{k_i+\frac{1}{2}}}{\sqrt{2}} \frac{k_i(2k_i+1)!!}{(k_i+1)!}\Big)\times
		\cH(k_1,k_2,k_3,k_4,\ldots)
\end{align}
where $\cH(k_1,k_2,k_3,k_4,\ldots)$ represents the total Wick contraction computed 
with the ``propagator''
\begin{align}
	\label{defHilb}
		H_{k_1,k_2} = \frac{1}{k_1 + k_2 + 1}~.
\end{align}
For instance, with 4 components one has
\begin{align}
	\label{wick4}
		\cH(k_1,k_2,k_3,k_4) = 
		H_{k_1,k_2} H_{k_3,k_4} + H_{k_1,k_3} H_{k_2,k_4} + H_{k_1,k_4} H_{k_2,k_3}~.
\end{align}
Similarly, with 6 components one has the sum of the 15 possible ways 
to make a complete contraction, and so on.

In order to exploit this Wick-like factorization property in the simplest way,
it is convenient to rephrase it in terms of normal-ordered and normalized operators. 

\paragraph{Normal-ordered operators:}
At the leading large-$N$ order the normal ordered version $O_n^{(0)}(a)$ of the single-trace 
operator $\tr a^n$ contains only single traces of lower powers of $a$ and is given by
\begin{align}
	\label{Oisp}
		O_n^{(0)}(a) = \tr p_n(a)
\end{align} 
where the monic polynomial $p_n(a)$ is related the Chebyshev polynomials of the first kind $T_n(x)$ as 
follows \cite{Rodriguez-Gomez:2016cem}:
\begin{align}
	\label{omega0isC2}
		p_n(a) = 2 \Big(\frac{N}{2}\Big)^{\frac{n}{2}}\, 
		T_n\Big(\frac{a}{\sqrt{2N}}\Big) + \delta_{n,2}\,\frac{N}{2}\,\mathbb{1} = a^n + \ldots~. 
\end{align}
For the first odd operators this amounts to
\begin{equation}
\begin{aligned}
		O^{(0)}_3(a)  & = \tr a^3~,\\
		O^{(0)}_5(a)  & = \tr a^5 - \frac{5}{2} N \tr a^3~,\\
		O^{(0)}_7(a)  & = \tr a^7 - \frac{7}{2} N \tr a^5  + \frac{7}{2} N^2 \tr a^3~,
\end{aligned}
	\label{noodd}
\end{equation}
while for the first even operators it gives
\begin{equation}
\begin{aligned}
		O^{(0)}_2(a)  & = \tr a^2-\frac{1}{2}N^2~,\\
		O^{(0)}_4(a)  & = \tr a^4 - 2N \tr a^2+\frac{1}{2}N^3~,\\
		O^{(0)}_6(a)  & = \tr a^6 - 3N \tr a^4  + \frac{9}{4} N^2 \tr a^2-\frac{1}{4}N^4~,
\end{aligned}
	\label{noeven}
\end{equation}
These normal-ordered operators have diagonal two-point functions:
\begin{align}
	\label{G0}
		\big\langle O_n^{(0)}(a)\,O_m^{(0)}(a)\big\rangle_{(0)} =
		G_n^{(0)}\,\delta_{n,m}~
		\qquad\text{with}~~~~~
		G_n^{(0)}	= n\Big(\frac{N}{2}\Big)^n~.		
\end{align}

\paragraph{The free-variable representation:}
Focusing on the odd sector, we introduce a normalized version of the normal-ordered 
operators, defining
\begin{align}
	\label{defome}
		\omega_k(a) = \frac{O^{(0)}_{2k+1}(a)}{\sqrt{G^{(0)}_{2k+1}}}~.
\end{align}
At large $N$, these objects have canonical two-point functions:
\begin{align}
	\label{corrome}
		\big\langle \omega_k(a)\, \omega_\ell(a)\big\rangle_{0} = \delta_{k\ell}~.
\end{align}
The linear relation between the $\omega_k(a)$'s and the odd single traces ensures that, at the leading order for large $N$, the correlators of many $\omega$'s are computed using Wick's theorem 
just as it was the case for the correlators of many odd single traces 
(see (\ref{tracewick})--(\ref{wick4})). 
Therefore, we can associate to each operator $\omega_k(a)$ a real variable
$\omega_k$ and write 
\begin{align}
	\label{corrome1}
		\big\langle \omega_{k_1}(a)\, \omega_{k_2}(a) \ldots \omega_{k_n}(a)\big\rangle_{0} 
		= \int \!D\boldsymbol{\omega}\, \,\omega_{k_1}\,\omega_{k_2}\ldots \omega_{k_n} \, 
		\rme^{-\frac 12 \,\boldsymbol{\omega}^T \,\boldsymbol{\omega}}~,
\end{align}		
where $\boldsymbol{\omega}$ is the (infinite) column vector of components $\omega_k$ and
\begin{align}
	\label{defDome}
		D\boldsymbol{\omega} = \prod_{k=1}^\infty \!\frac{d \omega_k}{\sqrt{2\pi}}~.
\end{align}  

In this representation it is quite easy to include the interactions of the $\mathbf{E}$ matrix model,
which are an infinite sum of double odd traces. By re-expressing these traces 
in terms of the operators $\omega_k(a)$, the interaction action (\ref{Sint}) takes the following form
\begin{align}
	\label{Soddome}
		S_{\mathrm{int}}(a)= - \frac{1}{2}\, \boldsymbol{\omega}^T\,\mathsf{X}\,\,\boldsymbol{\omega} ~,
\end{align}
where $\Xx$ is an infinite symmetric matrix depending on $\lambda$ originally introduced in \cite{Beccaria:2020hgy}.
The entries of $\Xx$ are initially expressed as series in $\lambda$ since this is how they are generated but, remarkably, these series can be resummed in terms of an integral of Bessel 
functions of the first kind, according to \cite{Beccaria:2020hgy}
\begin{align}
	\label{Xx}
		\mathsf{X}_{k\ell} = -8 (-1)^{k+\ell} \sqrt{(2k+1)(2\ell+1)} \int_0^\infty \!\frac{dt}{t}\, 
		\frac{\rme^t}{(\rme^t-1)^2}\,
		J_{2k+1}\Big(\frac{t\sqrt{\lambda}}{2\pi}\Big)\, 
		J_{2\ell+1}\Big(\frac{t\sqrt{\lambda}}{2\pi}\Big)~.	
\end{align}
This expression can be easily re-expanded in series of $\lambda$ at very high order, thus making it possible to obtain very long perturbative series in an efficient way. Moreover, as we will discuss 
in Section \ref{sec:large}, the entries $\mathsf{X}_{k\ell} $ can also be expanded for large $\lambda$ using Mellin transform techniques. This is the key to access the strong coupling behavior of all observables that can be given in terms of $\Xx$. 

A prime example of such observables is provided by the partition function (\ref{ZS4})
which can be rewritten as
\begin{align}
	\label{partE}
		\cZ = \Big\langle\rme^{-S_{\rm int}(a)}\Big\rangle_{0} = \Big\langle
		\rme^{\,\frac 12 \,\boldsymbol{\omega}^T\,\mathsf{X}\,\,\boldsymbol{\omega}}\Big\rangle_{0}~. 
\end{align}
Since the Wick property exponentiates, so does the correspondence (\ref{corrome1}). Therefore, we can re-express $\cZ$ in the free-variable representation as
\begin{align}
	\label{Zfree}
		\cZ = 
		\int\!D\boldsymbol{\omega}
		\, \rme^{-\frac{1}{2} \,\boldsymbol{\omega}^T\, (\mathbb{1} - \mathsf{X})\, \boldsymbol{\omega}} 
		\,=\,\mathrm{det}^{-\frac{1}{2}}\big(\mathbb{1}-\mathsf{X}\big)~.
\end{align}  
The corresponding free energy $\cF$ is then given by
\begin{align}
	\label{Fen}
		\cF = -\log\cZ=\frac 12 \tr \log \big(\mathbb{1}-\mathsf{X}\big)~.
\end{align}
Note that this $\cF$ is actually the free energy in the ``difference theory'', namely
$\cF = F^{\cN=2}-F^{\cN=4}$.

\subsection{Twisted observables}
\label{subsec:twisted} 
The vacuum expectation value of any operator $f(\omega(a))$ that can be written in terms of $\omega_k(a)$ or, equivalently, purely in terms of the odd traces of $a$, can be realized using the
free variables as follows
\begin{align}
	\label{vevEome2}
		\big\langle f\big(\omega(a)\big)\big\rangle 
		= \frac{1}{\cZ}
		\int\!D\boldsymbol{\omega}\, f(\boldsymbol{\omega})\, \rme^{-\frac{1}{2} \,\boldsymbol{\omega}^T\, (\mathbb{1} - \mathsf{X})\, \boldsymbol{\omega}}~.
\end{align}
From this expression, we see that the operators $\omega_k(a)$, which were diagonal in the Gaussian theory, in the $\mathbf{E}$ model have the following propagator\,%
\footnote{Here and in the following, for the sake of visual clarity but with an abuse of notation, we will often indicate as ${1}/{M}$ the inverse of a matrix $M$.}
\begin{align}
	\label{omeome}
		\big\langle \omega_k(a)\,\omega_\ell(a) \big\rangle
		= \Big(\frac{1}{\mathbb{1} - \mathsf{X}}\Big)_{k,\ell}
		\equiv \Dx_{k,\ell} 
		~.
\end{align}
This propagator $\Dx$ is an infinite matrix with a highly non-trivial dependence on $\lambda$.
Multiple correlators $\big\langle \omega_{k_1}(a)\,\omega_{k_2}(a)\cdots \big\rangle$ can just be treated as tree-level expressions and computed by applying Wick's theorem using $\Dx$ as propagator. 

From these results it easy to deduce the expression of the two-point correlator of any two odd traces. In fact, it is sufficient to express these odd traces in terms of the $\omega_k(a)$'s by inverting the relations described in
(\ref{Oisp}), (\ref{omega0isC2}) and (\ref{defome}). 
Explicitly, with the help of Eq. (5.6) of \cite{Beccaria:2020hgy}, one gets
\begin{align}
	\label{trtoome}
		\tr a^{2k+1} = \Big(\frac{N}{2}\Big)^{k + \frac 12}\, \sum_{i=0}^{k-1}
		c_{k,i} \, \omega_{k-i}(a)~,
\end{align}
where the coefficients $c_{k,i}$ are
\begin{align}
	\label{cikis}
		c_{k,i} =\binom{2k+1}{i} \sqrt{2k-2i+1}~.
\end{align}
Then, the correlator of any two odd traces in the $\mathbf{E}$ model is given by 
\begin{align}
	\label{tpoddtr}
		\big\langle\tr a^{2k+1}\tr a^{2\ell+1}\big\rangle 
		= \Big(\frac{N}{2}\Big)^{k + \ell+ 1} \,\sum_{i=0}^{k-1} \sum_{j=0}^{\ell-1}
		c_{k,i} c_{\ell,j}\, \Dx_{k-i,\ell-j}~.  
\end{align}
A few explicit examples are:
\begin{subequations}
\begin{align}
\big\langle\tr a^{3}\tr a^{3}\big\rangle &=\frac{3N^3}{8}\,\Dx_{1,1}~,\\[1mm]
\big\langle\tr a^{3}\tr a^{5}\big\rangle &=\frac{15N^4}{16}\,\Big(\Dx_{1,1}+\frac{1}{\sqrt{15}}
\,\Dx_{1,2}\Big)~,\\[1mm]
\big\langle\tr a^{5}\tr a^{5}\big\rangle &=\frac{5N^5}{2}\,\Big(\frac{15}{16}\,\Dx_{1,1}+
\frac{\sqrt{15}}{8}\,\Dx_{1,2}+\frac{1}{16}\,\Dx_{2,2}\Big)~.
\end{align}
\label{oddodd}%
\end{subequations}
Since each propagator $\Dx_{k,\ell}$ is known \emph{exactly} in terms of $\lambda$ through 
the matrix $\Xx$, the above expressions represent a complete resummation of the perturbative
results that have been previously considered in the literature. 
In turn, each $\Dx_{k,\ell}$ can be re-expanded in
powers of $\lambda$ and used to generate in an efficient way the perturbative series.
This expansion can be done in two steps: firstly, by expanding in powers of 
$\Xx$, then by expanding the matrix $\Xx$ using its definition (\ref{Xx}). Thus, we first write
\begin{equation}
\Dx_{k,\ell}=\delta_{k,\ell}+\Xx_{k,\ell}+\Xx^2_{k,\ell}+\Xx^3_{k,\ell}+\ldots~,
\label{Dexp}
\end{equation}
and then we use the sum rule
\begin{equation}
\cG(t,t^\prime)=8\sum_{k=1}^\infty(2k+1)J_{2k+1}(t)J_{2k+1}(t^\prime)
=-\frac{4\, t t^\prime}{t^2-{t^\prime}^2}\Big[t\, J_1(t)J_2(t^\prime)-t^\prime\,
J_2(t)J_1(t^\prime)\Big]
\label{sumrule}
\end{equation}
to obtain
\begin{align}
\Xx^2_{k,\ell}&=8 (-1)^{k+\ell} \sqrt{(2k+1)(2\ell+1)} \int\!\cD t\,\cD t^\prime ~J_{2k+1}(zt)\,
\cG(zt,zt^\prime)\,J_{2\ell+1}(zt^\prime)~,
\label{Xx23}\\
\Xx^3_{k,\ell}&=-8 (-1)^{k+\ell} \sqrt{(2k+1)(2\ell+1)} \int\!\cD t\,\cD t^\prime\,\cD t^{\prime\prime} ~J_{2k+1}(zt)\,
\cG(zt,zt^\prime)\,\cG(zt^\prime,zt^{\prime\prime})\,J_{2\ell+1}(zt^{\prime\prime})~,
\notag
\end{align}
and so on, where we have defined for convenience
\begin{equation}
\cD t=\frac{dt}{t}\, 
		\frac{\rme^t}{(\rme^t-1)^2}\quad\mbox{and}\quad
		z=\frac{\sqrt{\lambda}}{2\pi}~.
\end{equation}
These multiple integrals can be easily computed using the expansions of the Bessel functions.
In this way one finds that $\Xx^n_{k,\ell}$ contains $n$ powers of Riemann $\zeta$-values with odd arguments and corresponds to the resummation of all perturbative contributions that in the gauge theory arise from diagrams involving $n$ loops of matter hypermultiplets in the planar limit.

\paragraph{Normal-ordering in the $\mathbf{E}$ theory:}
While the above exact results for correlators in the matrix model are interesting by themselves, our main interest is in their bearing on the $\cN=2$ SYM field theory of type $\mathbf{E}$ in flat space.
The first step we have to carry out to reach this goal, is the Gram-Schmidt procedure to obtain the normal-ordered operators 
\begin{equation}
\widehat\omega_k(a)
\end{equation}
that represent, at the leading large-$N$ order, the chiral operators $\tr \varphi^{2k+1}(x)$ 
of the $\cN=2$ SYM theory, rescaled so as to have unit correlators in the Gaussian model according to (\ref{corrome}).
In other words, the $\widehat\omega_k(a)$'s are related to the operators $O_{2k+1}(a)$ 
introduced in Section~\ref{secn:matrix} by
\begin{align}
	\label{wotoO}
		\widehat\omega_k(a) = \frac{O_{2k+1}(a)}{\sqrt{G^{(0)}_{2k+1}}}~.
\end{align}
By definition, these normal-ordered operators have diagonal two-point functions, which we write as
\begin{align}
	\label{fiome}
		 \big\langle
		 \widehat\omega_k(a)\, \widehat\omega_\ell(a)\big\rangle = \big(1 + \Delta_k(\lambda)\big) \delta_{k\ell}~,
\end{align}
where $\Delta_k(\lambda)$ vanishes for $\lambda\to 0$. These operators capture precisely the twisted observables of the gauge theory defined in (\ref{Deltaodd}), namely
\begin{align}	
	\label{f_i}
\frac{\big\langle
		\tr \varphi^{2k+1}(x)\,\tr \overbar\varphi^{2k+1}(0)\big\rangle\phantom{\Big|}}
		{\big\langle
		\tr \varphi^{2k+1}(x)\,\tr \overbar\varphi^{2k+1}(0)\big\rangle_0\phantom{\Big|}}
		= 1 + \Delta_k(\lambda)+ O(N^{-2})~.
\end{align}       

The key observation is that the Gram-Schmidt procedure expresses the two-point function of the normal ordered operators $\widehat\omega_k(a)$ as a ratio of determinants of the correlation matrix (\ref{omeome}), so that 
\begin{align}
	\label{gammaome0}
	1 + \Delta_k(\lambda) = 
		 \frac{\det \Dx_{(k)}}{\det \Dx_{(k-1)}}~. 
\end{align} 		
Here we have introduced the notation $M_{(k)}$ to indicate
the upper left $k\times k$ block of a matrix $M$, with the convention that $M_{(0)}= 1$. As shown 
in \cite{Beccaria:2020hgy}, the ratio of determinants in (\ref{gammaome0}) can also be rewritten in a different and more suggestive way. To do so let us define $M_{[k]}$ as the sub-matrix which is obtained from $M$ by removing its first $(k-1)$ rows and columns; in this way $M_{[1]} = M$. Then, using the definition (\ref{omeome}), one can prove that for any $k$
\begin{align}
	\label{gammaome}
1 + \Delta_k(\lambda) =
		\bigg(\frac{1}{\mathbb{1} - \Xx_{[k]}}\bigg)_{1,1} ~.
\end{align}
We stress once again that this result is exact in $\lambda$.

At weak coupling, it is quite straightforward to extract the perturbative expansion from the above expression. Indeed, one first expands (\ref{gammaome}) in powers of $\Xx_{[k]}$ to get
\begin{equation}
\Delta_k(\lambda) =\big(\Xx_{[k]}\big)_{1,1}+\big(\Xx^2_{[k]}\big)_{1,1}+\big(\Xx^3_{[k]}\big)_{1,1}
+\ldots
\end{equation}
where
\begin{subequations}
\begin{align}
\big(\Xx_{[k]}\big)_{1,1}&=\Xx_{k,k}~,\\
\big(\Xx^2_{[k]}\big)_{1,1}&=\Xx^2_{k,k}-\sum_{m=1}^{k-1}\Xx_{k,m}\Xx_{m,k}~,\\
\big(\Xx^3_{[k]}\big)_{1,1}&=\Xx^3_{k,k}-2\sum_{m=1}^{k-1}\Xx_{k,m}\Xx^2_{m,k}+
\sum_{m,n=1}^{k-1}\Xx_{k,m}\Xx_{m,n}\Xx_{n,k}~,
\end{align}%
\end{subequations}
and so on, and then uses the sum rule (\ref{sumrule}) to reduce the calculation to multiple integrals
of Bessel functions. This procedure can be easily automatized and the perturbative expansion
of $\Delta_k(\lambda)$ can be obtained to very high orders with relatively short computational effort. For instance, at the first few orders for $k=1$ and $k=2$, one gets
\begin{align}
		\Delta_1(\lambda) & =-\frac{5\,\zeta (5)}{256\pi^6}\lambda^3+\frac{105\, \zeta (7)}{4096\pi^8}\lambda^4-\frac{1701\, \zeta (9)}{65536\pi^{10}} \lambda^5+
		\Big(\frac{25\,\zeta (5)^2}{65536\pi^{12}}+\frac{12705\, \zeta (11)}{524288\pi^{12}}\Big)\lambda^6\notag\\[1mm]
		&\quad-\Big(\frac{525 \,\zeta (5) \zeta (7)}{524288\pi^{14}}+\frac{184041\, \zeta (13)}{8388608\pi^{14}}\Big)\lambda^7+\ldots~,
		\label{Delta1}\\[3mm]
		\Delta_2(\lambda) & = -\frac{63 \,\zeta(9)}{65536 \pi^{10}}\lambda^5 
		+\frac{1155\,\zeta (11)}{524288 \pi^{12}}\lambda^6 -\frac{27885\, \zeta(13)}{8388608 
		\pi^{14}}\lambda^7 +\ldots~. \label{Delta2}
\end{align}
We have actually worked out these expansions up to order $\lambda^{160}$, and constructed from them Pad\'e approximants that turn out to be extremely stable far beyond the convergence radius of the perturbative expansion, as we will see in Section~\ref{sec:numerical}.

\subsection{Untwisted observables}
\label{subsec:untwisted}

Let us now turn to the ``untwisted'' observables and consider an even-trace operator $\tr a^{2k}$ in the matrix model. Writing its expectation value as
\begin{equation}
\label{t2k}
\big\langle \tr a^{2k} \big\rangle= \frac{\big\langle \tr a^{2k}\,\, \rme^{-S_{\rm int}(a)}\big\rangle_0\phantom{\Big|}}{\big\langle \rme^{-S_{\rm int}(a)}\big\rangle_0\phantom{\Big|}}
=\big\langle \tr a^{2k} \big\rangle_0-\cA_{2k}~,
\end{equation}
we realize that, at the first order in the interaction action, the deviation $\cA_{2k}$ from the 
Gaussian result is given by
\begin{align}
	\label{t2nsint}
\cA_{2k}=\big\langle \tr a^{2k} \,S_{\rm int}(a)\big\rangle_0 - 
\big\langle \tr a^{2k} \big\rangle_0 \,\big\langle S_{\rm int}(a)\big\rangle_0 ~.
\end{align} 
As shown in Section~\ref{secn:matrix}, the interaction action $S_{\rm int}(a)$ is a collection of double odd traces and
can be recast in the form
\begin{align}
	\label{Sintodd}
		S_{\mathrm{int}}(a) = \sum_{p=2}^\infty \,\sum_{q=1}^{p-1}
		\Big(\frac{\lambda}{8\pi^2 N}\Big)^{p+1} f_{p,q} \, \tr a^{2q+1}\, \tr a^{2p-2q+1}
\end{align}
where the coefficients $f_{p,q}$ can be easily read from (\ref{Sint}). 
Then, using the same notation as in (\ref{toddis}),
the correction (\ref{t2nsint}) reads
\begin{align}
	\label{S2nis}
		\mathcal{A}_{2k} & = \sum_{p=2}^\infty
		\,\sum_{q=1}^{p-1}  \Big(\frac{\lambda}{8\pi^2 N}\Big)^{p+1}  f_{p,q}\,\Big[
		t_{2k,2q+1,2p-2q+1} -t_{2k}\,t_{2q+1,2p-2q+1}\Big]~.
\end{align}
As proved in \cite{Beccaria:2020hgy}, the even traces factorize in every Gaussian correlator at the planar level, meaning that
\begin{equation}
t_{2k,n,m,\ldots}=t_{2k}\,t_{n,m,\ldots}
\label{factor}
\end{equation}
at the planar level.
Therefore, the first contribution to the connected correlator appearing in 
$\mathcal{A}_{2k}$ occurs at the sub-leading order and takes the form
\begin{align}
	\label{conn2n2odd}
	t_{2k,2q+1,2p-2q+1}-t_{2k}\,t_{2q+1,2p-2q+1}=\frac{k(k+1)}{N^2} \, (p+1)\, 
	t_{2k}\,t_{2q+1,2p-2q+1} + \ldots		 
\end{align}
where the ellipses denote terms suppressed at large $N$. This contribution arises when just two propagators connect the even trace to the odd ones. Inserting this result into (\ref{S2nis}), we
get
\begin{equation}
\begin{aligned}
	\label{S2nis2}
		\mathcal{A}_{2k} & = \frac{k(k+1)}{N^2}\, t_{2k}\,
		\sum_{p=2}^\infty \,\sum_{q=1}^{p-1}\,
		(p+1) \Big(\frac{\lambda}{8\pi^2 N}\Big)^{p+1} f_{p,q}\,\,t_{2q+1,2p-2k+1}+\ldots
		\\[1mm]
		& = \frac{k(k+1)}{N^2} \,t_{2k}~\lambda\, \partial_\lambda
		\big\langle S_{\mathrm{int}}(a)\big\rangle_0+\ldots~,
\end{aligned}	
\end{equation}
so that, at the lowest order in $S_{\mathrm{int}}(a)$, we can rewrite (\ref{t2k}) as follows
\begin{align}
	\label{t2nratio}
		\frac{\big\langle \tr a^{2k}\big\rangle\phantom{\Big|}}{\big\langle \tr a^{2k}\big\rangle_0
		\phantom{\Big|}} =1 - \frac{k(k+1)}{N^2}\,\, \lambda \,\partial_\lambda\vev{S_{\rm int}}_0 + \ldots~.
\end{align}
The same pattern persists also when we consider higher powers of $S_{\mathrm{int}}(a)$ and at order $1/N^2$ we get contributions from connected diagrams where the even trace is linked to the odd ones. Thus we can express the result as
\begin{align}
	\label{t2nratiofull}
		\frac{\big\langle \tr a^{2k}\big\rangle\phantom{\Big|}}{\big\langle \tr a^{2k}\big\rangle_0
		\phantom{\Big|}} 
		= 1 + \frac{k(k+1)}{N^2}\, \frac{\lambda \,\partial_\lambda
		\big\langle \rme^{-S_{\mathrm{int}}(a)}\big\rangle_0\phantom{\Big|}}{\big\langle
		\rme^{-S_{\mathrm{int}}(a)}\big\rangle_0\phantom{\Big|}} + \ldots
		= 1 - \frac{k(k+1)}{N^2} \lambda \,\partial_\lambda \cF+ \ldots ~.
\end{align}
where $\cF$ is the free energy (\ref{Fen}).

If we consider multiple even traces, the contribution of order $1/N^2$ to their vacuum expectation
value arises when just one of them is linked to the odd traces coming from the interaction action. Thus, we simply have to sum over all such possibilities, getting
\begin{align}
	\label{t2niratiofull}
		\frac{\big\langle \tr a^{2k_1}\,\tr a^{2k_2}\cdots
		\big\rangle\phantom{\Big|}}{\big\langle \tr a^{2k_1}\,
		\tr a^{2k_2}\cdots\big\rangle_0
		\phantom{\Big|}} 
		 = 1 - \frac{\sum_l \,k_l(k_l+1)}{N^2} \,
		 \lambda\, \partial_\lambda \cF+ \ldots ~.
\end{align}
Since the free energy $\cF$ is known in terms of the matrix $\Xx$ as indicated in (\ref{Fen}),
also all these vacuum expectation values can be expressed in terms of $\Xx$ and
thus the full $\lambda$-dependence of their first non-planar corrections is completely determined by (\ref{t2niratiofull}).

A similar reasoning can be used to find the correlators of the normal-ordered
operators $O_{2k}(a)$ that represent the single-trace observables of even weight in the gauge theory. To reach this goal, we have first to determine how $O_{2k}(a)$ is written in terms of the even traces
by means of the Gram-Schmidt diagonalization method. Once this step is completed, we can compute
the correlators and see how they behave. This procedure is completely straightforward but rather tedious. Therefore, here we discuss only the simplest case corresponding 
to $k=1$. In this case the normal-ordered operator associated to
the chiral operator $\tr \varphi(x)^2$ is just
\begin{equation}
O_2(a)=\tr a^2-\big\langle \tr a^2\big\rangle~,
\label{O2is}
\end{equation}
and thus its two-point correlator is
\begin{equation}
\big\langle O_2(a)\,O_2(a)\big\rangle =\big\langle \tr a^2\, \tr a^2 \big\rangle
-\big\langle \tr a^2\big\rangle^2=
\big\langle \tr a^2\, \tr a^2 \big\rangle_0
-\big\langle \tr a^2\big\rangle_0^2-\cA_{2,2}~.
\label{O2O2is}
\end{equation}
At the first order in the interaction action, the deviation from the Gaussian result is
\begin{equation}
\begin{aligned}
\cA_{2,2}&=\big\langle \tr a^2\, \tr a^2 \,S_{\mathrm{int}}(a)\big\rangle_0
-\big\langle \tr a^2\, \tr a^2 \big\rangle_0\,\big\langle S_{\mathrm{int}}(a)\big\rangle_0\\
&\quad-2\big\langle \tr a^2 \big\rangle_0\,\big\langle \tr a^2 \,S_{\mathrm{int}}(a)\big\rangle_0
+2\big\langle \tr a^2 \big\rangle_0^2\,\big\langle S_{\mathrm{int}}(a)\big\rangle_0~.
\end{aligned}
\end{equation}
Using the form of the interaction action given in (\ref{Sintodd}), we can rewrite the previous
expression as
\begin{equation}
\begin{aligned}
\cA_{2,2}&= \sum_{p=2}^\infty
		\,\sum_{q=1}^{p-1}  \Big(\frac{\lambda}{8\pi^2 N}\Big)^{p+1}  f_{p,q}\,\Big[
		t_{2,2,2q+1,2p-2q+1}-t_{2,2}\,t_{2q+1,2p-2q+1}\\
		&\qquad\qquad\qquad\qquad
		-2\, t_{2}\,t_{2,2q+1,2p-2q+1}+2 \,t_{2}^2\,t_{2q+1,2p-2q+1}\Big]~.
		\end{aligned}
\end{equation}
The quantity in square brackets can be evaluated by exploiting the relations
\begin{equation}
t_2=\frac{N^2-1}{2}~,\quad
t_{2,n,m,\ldots}=\frac{N^2-1+n+m+\ldots}{2}\,t_{n,m,\ldots}~,
\label{rel}
\end{equation}
which were derived in \cite{Billo:2017glv} using the fusion/fission identities of the traces of SU($N$).
In this way, after straightforward algebra, we find 
\begin{equation}
\begin{aligned}
\cA_{2,2}&= \sum_{p=2}^\infty
		\,\sum_{q=1}^{p-1} \, (p+1)(p+2)\,\Big(\frac{\lambda}{8\pi^2 N}\Big)^{p+1}  f_{p,q}\,
		t_{2q+1,2p-2q+1}=\Big(\lambda\,\partial_\lambda+\big(\lambda\,\partial_\lambda)^2\Big)
		\big\langle S_{\mathrm{int}}(a)\big\rangle_0~.
\end{aligned}
\label{A22expl}
\end{equation}
Finally, taking into account that at large $N$
\begin{equation}
\big\langle O^{(0)}_2(a)\,O^{(0)}_2(a)\big\rangle_0=
\big\langle \tr a^2\, \tr a^2 \big\rangle_0
-\big\langle \tr a^2\big\rangle_0^2=t_{2,2}-t_2^2=\frac{N^2}{2}+\ldots~,
\end{equation}
we deduce from (\ref{O2O2is}) and (\ref{A22expl}) that
\begin{equation}
\frac{\big\langle O_2(a)\,O_2(a)\big\rangle\phantom{\Big|}}{\big\langle O^{(0)}_2(a)\,O^{(0)}_2(a)\big\rangle_0\phantom{\Big|}}=1+\frac{1}{N^2}\,\Big(2\lambda\,\partial_\lambda+2\big(\lambda\,\partial_\lambda)^2\Big)
		\big\langle S_{\mathrm{int}}(a)\big\rangle_0+\ldots
\end{equation}
where the ellipses stand for terms suppressed at large $N$.

This structure persists at higher orders in $S_{\mathrm{int}}(a)$ and, as before, it exponentiates allowing us to express the final result in terms of the free energy $\cF$ in the following way
\begin{equation}
\frac{\big\langle O_2(a)\,O_2(a)\big\rangle\phantom{\Big|}}{\big\langle O^{(0)}_2(a)\,O^{(0)}_2(a)\big\rangle_0\phantom{\Big|}}=1-\frac{1}{N^2}\,\Big(2\,\lambda\,\partial_\lambda\cF+2\big(\lambda\,\partial_\lambda)^2\cF\Big)+\ldots~.
\end{equation}
Mapping this to the chiral/anti-chiral correlator of the gauge theory, we find precisely
the form anticipated in (\ref{Deltaeven}), namely
\begin{equation}
\frac{\big\langle
		\tr \varphi^{2}(x)\,\tr \overbar\varphi^{2}(0)\big\rangle\phantom{\Big|}}
		{\big\langle
		\tr \varphi^{2}(x)\,\tr \overbar\varphi^{2}(0)\big\rangle_0\phantom{\Big|}}
		= 1 + \frac{\delta_1(\lambda)}{N^2}+ O(N^{-4})~,
		\label{O2O2exact}
\end{equation}
with
\begin{equation}
\delta_1(\lambda)=-2\,\Big[\lambda\,\partial_\lambda\cF+\big(\lambda\,\partial_\lambda)^2\cF\Big]~.
\label{delta1}
\end{equation}
We have also analyzed in this way the cases $k=2$ and $k=3$, and found
\begin{equation}
\begin{aligned}
\delta_2(\lambda)&=-4\,\Big[7\,\lambda\,\partial_\lambda\cF+\big(\lambda\,\partial_\lambda)^2\cF\Big]~,\\
\delta_3(\lambda)&=-6\,\Big[17\,\lambda\,\partial_\lambda\cF+\big(\lambda\,\partial_\lambda)^2\cF\Big]
~,
\end{aligned}
\label{delta2}
\end{equation}
leading us to conjecture that
\begin{equation}
\delta_k(\lambda)=-2k\,\Big[(2k^2-1)\,\lambda\,\partial_\lambda\cF+\big(\lambda\,\partial_\lambda)^2\cF\Big]~.
\label{deltak}
\end{equation}

These expressions contain the exact dependence on $\lambda$ since the free
energy $\cF$ can be written in closed form in terms of the matrix $\Xx$ as we
have seen in (\ref{Fen}). If one is interested in the perturbative expansion for
$\lambda\to 0$, this can be efficiently generated using
\begin{equation}
\cF=\frac 12 \tr \log \big(\mathbb{1}-\mathsf{X}\big)=-\frac{1}{2}\,\tr \Xx
-\frac{1}{4}\,\tr \Xx^2-\frac{1}{6}\,\tr \Xx^3+\cdots
\end{equation} 
and computing the traces using the sum rule (\ref{sumrule}) in (\ref{Xx}) and (\ref{Xx23}), that yields
\begin{equation}
\begin{aligned}
\tr \Xx &=-\int\!\cD t~\cG(zt,zt)~,\\
\tr \Xx^2&=\int\!\cD t\,\cD t^\prime ~
\cG(zt,zt^\prime)\,\cG(zt^\prime,zt)~,\\
\tr \Xx^3&=-\int\!\cD t\,\cD t^\prime\,\cD t^{\prime\prime} 
\cG(zt,zt^\prime)\,\cG(zt^\prime,zt^{\prime\prime})\,\cG(zt^{\prime\prime},zt)~,
\end{aligned}
\end{equation}
and so on. Evaluating these multiple integrals after expanding the Bessel functions
for small $\lambda$, we have found at the first few perturbative orders that
\begin{align}
		\delta_1(\lambda) & =-\frac{15\,\zeta (5)}{64\pi^6}\lambda^3+\frac{525\, \zeta (7)}{1024\pi^8}\lambda^4-\frac{6615\, \zeta (9)}{8192\pi^{10}} \lambda^5+
		\Big(\frac{525\,\zeta (5)^2}{65536\pi^{12}}+\frac{72765\, \zeta (11)}{65536\pi^{12}}\Big)\lambda^6\notag\\[1mm]
		&\quad-\Big(\frac{3675 \,\zeta (5) \zeta (7)}{131072\pi^{14}}+\frac{1486485\, \zeta (13)}{1048576\pi^{14}}\Big)\lambda^7+\ldots~,
		\label{delta1exp}\\[3mm]
		\delta_2(\lambda) & =-\frac{75\,\zeta (5)}{64\pi^6}\lambda^3+\frac{1155\, \zeta (7)}{512\pi^8}\lambda^4-\frac{6615\, \zeta (9)}{2048\pi^{10}} \lambda^5+
		\Big(\frac{975\,\zeta (5)^2}{32768\pi^{12}}+\frac{135135\, \zeta (11)}{32768\pi^{12}}\Big)\lambda^6\notag\\[1mm]
		&\quad-\Big(\frac{25725 \,\zeta (5) \zeta (7)}{262144\pi^{14}}+\frac{10405395\, \zeta (13)}{2097152\pi^{14}}\Big)\lambda^7+\ldots~. \label{delta2exp}
\end{align}
These expansions can be easily pushed to very high orders without too much computational effort  and then used for numerical investigations
(see also \cite{Beccaria:2021vuc} for similar calculations
related to the vacuum expectation value of the BPS Wilson loop).

\section{Analytic results at strong coupling}
\label{sec:large}
The strong coupling behavior of the observables computed so far can be studied by investigating the expansion for $\lambda\to\infty$ 
of the invariants related to the matrix $\Xx$.
The form of $\mathsf{X}$ at large $\lambda$ can be obtained 
by writing the products of the Bessel functions appearing in its definition as an inverse 
Mellin transform:
\begin{equation}
\label{jj}
\begin{aligned}
J_{2k+1}\Big(\frac{t\sqrt{\lambda}}{2\pi}\Big)\, 
& J_{2\ell+1}\Big(\frac{t\sqrt{\lambda}}{2\pi}\Big)\\[2mm]
= &  
\int_{- {\ii} \infty}^{ +{\ii} \infty} \, \frac{ds}{2 \pi \ii}\,
\frac{\Gamma(-s)\, \Gamma(2s+2k+2\ell+3)}{\Gamma(s+2k+2)\Gamma(s+2\ell+2)\Gamma(s+2k+2\ell+3)} \,
 \Big(\frac{t\sqrt{\lambda}}{4\pi}\Big)^{2(s+k+\ell+1)}~.
\end{aligned}
\end{equation}
Using the identity 
\begin{align}
\label{zet}
\int_0^\infty dt\, \frac{\rme^t}{(\rme^t-1)^2}\, t^{2n+1} = 
\Gamma (2n+2) \,\zeta(2n+1) ~,
\end{align}
the matrix elements $\mathsf{X}_{k\ell}$ given (\ref{Xx}) can be written as
\begin{equation}
\label{mtx}
\begin{aligned}
\mathsf{X}_{k\ell}
&= - 8 (-1)^{k+\ell} \sqrt{(2k+1)(2\ell+1)} \,
\int_{- \ii\infty}^{ +\ii\infty} \, \frac{ds}{2 \pi \ii}\,\bigg[
\Big(\frac{\sqrt{\lambda}}{4\pi}\Big)^{2(s+k+\ell+1)} \\[2mm]
& \quad\quad\times 
\frac{\Gamma(-s)\, \Gamma(2s+2k+2\ell+3) \,\Gamma(2s+2k+2\ell+2)}
{\Gamma(s+2k+2)\Gamma(s+2\ell+2)\Gamma(s+2k+2\ell+3)}
\,\zeta(2s+2k+2\ell+1)\bigg]~.
\end{aligned}
\end{equation}
The asymptotic expansion of this expression for large $\lambda$ receives contributions from the poles on the negative real axis and reads
\begin{equation}
\label{mtrx}
\begin{aligned}
\mathsf{X}_{k\ell}
& = - 8 (-1)^{k+\ell} \sqrt{(2k+1)(2\ell+1)} \,
\bigg[ \frac{\lambda}{16 \pi^2} \Big( \frac{\delta_{k-1,\ell}}{2(2k-1)\, 2k \,(2k+1)} +
\frac{\delta_{k,\ell}}{2k\,(2k+1)\,(2k+2)} \\[1mm]
& \quad + \frac{\delta_{k+1,\ell}}{2(2k+1)\,(2k+2)\,(2k+3)} \Big) 
- \frac{\delta_{k,\ell}}{24 (2k+1)} + O(\lambda^{-1/2}) \bigg]~.
\end{aligned}
\end{equation}
Therefore, at strong coupling one has
\begin{align}
	\label{xs}
		\mathsf{X} \underset{\lambda \to \infty}{\sim}
		- \frac{\lambda}{2 \pi^2}  \, \Sx ~,
\end{align}
where $\Sx$ is a three-diagonal infinite matrix whose elements are \cite{Beccaria:2021vuc}
\begin{align}
\label{s}
\mathsf{S}_{k\ell}
=  \frac{1}{4} (-1)^{k+\ell} \sqrt{\frac{2\ell+1}{2k+1}} \,
\Big(\frac{\delta_{k-1,\ell}}{k\,(2k-1)} + \frac{\delta_{k,\ell}}{k\,(k+1)} +
\frac{\delta_{k+1,\ell}}{(k+1)\,(2k+3)}\Big) ~.
\end{align}
It is quite remarkable that, as observed in \cite{Beccaria:2021vuc}, this is essentially the same 
matrix\,%
\footnote{Up to signs and an overall factor of $1/2$.} appearing in the analysis of the zeros
of Bessel functions performed in \cite{Ikebe}. Note that the main diagonal and the super- and 
sub-diagonals of $\mathsf{S}$ converge to 0 when the size of the matrix increases. This is the
key property that allows us to obtain finite results from this infinite matrix.

\subsection{Twisted observables}
In Section~\ref{subsec:twisted} we managed to express various twisted observables in terms of the 
propagator $\Dx$ of the free-variable representation. Using (\ref{xs}), we see that
at strong coupling
\begin{align}
	\label{Dll}
		\Dx = \Big(\frac{1}{\mathbb{1}-\Xx}\Big) \underset{\lambda \to \infty}{\sim}
		\frac{2\pi^2}{\lambda} \Sx^{-1}~,
\end{align} 
so that for large $\lambda$ the twisted observables scale as $1/\lambda$ at the leading order.

In Appendix \ref{app:details}, we show that it is possible to find an explicit form for the inverse 
of the infinite matrix $\Sx$. In particular, from (\ref{ymuno2app}) we obtain 
\begin{equation}
	\label{ymuno1}
		\Dx_{k\ell} \underset{\lambda \to \infty}{\sim} 
		\frac{4\pi^2}{\lambda} \sqrt{(2k+1)(2\ell+1)} \times
		\begin{cases}
			 k\,(k+1) & \text{for} ~k\leq \ell~,\\[2mm] 
			 \ell\,(\ell+1) & \text{for} ~k \geq \ell~. 
		\end{cases}
\end{equation}
This, together with (\ref{tpoddtr}), allows us to find the numerical coefficients in front 
of $1/\lambda$ in an explicit way for all the two-point correlators of the odd-trace operators.
For instance, in the examples considered in (\ref{oddodd}) we get 
\begin{subequations}
\begin{align}
		&\big\langle\tr a^{3}\tr a^{3}\big\rangle  \underset{\lambda \to \infty}{\sim} \frac{3N^3}{8}\times \frac{24\pi^2}{\lambda}+\ldots~,\\[1mm]
		&\big\langle\tr a^{3}\tr a^{5}\big\rangle \underset{\lambda \to \infty}{\sim} \frac{15N^4}{16}\times \frac{32\pi^2}{\lambda}+\ldots~,\\[1mm]
		&\big\langle\tr a^{5}\tr a^{5}\big\rangle \underset{\lambda \to \infty}{\sim} \frac{5N^5}{2}\times \frac{45\pi^2}{\lambda}+\ldots~.
\end{align}
\label{oddoddll}%
\end{subequations}
 
Most importantly, also the quantities $\Delta_k(\lambda)$ of the $\mathbf{E}$ gauge theory 
admit an exact expression in terms of the matrix $\Xx$, as we have shown in (\ref{gammaome0}) and (\ref{gammaome}), which we rewrite here for convenience:
\begin{align}
	\label{detDdetD}
	1 + \Delta_k(\lambda) = \frac{\det\Dx_{(k)}}{\det\Dx_{(k-1)}} = 
		\Big(\frac{1}{\mathbb{1}-\Xx_{[k]}}\Big)_{1,1}~.		
\end{align}
For any given $k$ we can use the expression in the middle, which is a ratio of determinants of finite matrices whose leading large-$\lambda$ behavior can be deduced from (\ref{ymuno1}). For example, for the lowest values of $k$ we have  
\begin{align}
	\label{Dlowk}
		D_{(0)} = 1~,~~~
		D_{(1)} \underset{\lambda \to \infty}{\sim}  \frac{4\pi^2}{\lambda} \, 6~,~~~
		D_{(2)} \underset{\lambda \to \infty}{\sim}  \frac{4\pi^2}{\lambda} \,
		\begin{pmatrix}
			6 & 2\sqrt{15} \\
			2 \sqrt{15} & 30
		\end{pmatrix}~.
\end{align}
Inserting this into the first equality of (\ref{detDdetD}), it immediately follows that
\begin{align}
	\label{D3D5}
		1 + \Delta_1(\lambda) \underset{\lambda \to \infty}{\sim}  \frac{24\pi^2}{\lambda}~,~~~
		1 + \Delta_2(\lambda) \underset{\lambda \to \infty}{\sim}  \frac{80\pi^2}{\lambda}~.
\end{align}

The second equality in (\ref{detDdetD}) allows to to obtain the result in closed form for generic $k$. 
In fact, at large $\lambda$ we can use (\ref{xs}) so that
\begin{align}
	\label{detDdetS}
		1 + \Delta_k(\lambda) \underset{\lambda \to \infty}{\sim} 
		 \frac{2\pi^2}{\lambda} \,\big(\,\Sx_{[k]}^{-1}\big)_{1,1}
		= \frac{2\pi^2}{\lambda} \,\frac{\det \Sx_{[k+1]}}{\det \Sx_{[k]}}~.
\end{align}
Note that in the last expression the ratio of the determinants of two infinite matrices appears. 
However, this ratio can be explicitly computed in closed form for any $k$, as we show in appendix \ref{app:details}. In the end, using (\ref{yn11}), we obtain
\begin{align}
	\label{resDeltak}
	1 + \Delta_k(\lambda) \underset{\lambda \to \infty}{\sim} 
		 \frac{8\pi^2}{\lambda}\, k(2k+1)~.
\end{align}
Therefore, from this analysis we find that the strong-coupling behavior of the two-point functions of chiral/anti-chiral operators
in the $\mathbf{E}$ superconformal gauge theory is given by
\begin{equation}
\frac{\big\langle
		\tr \varphi^{2k+1}(x)\,\tr \overbar\varphi^{2k+1}(0)\big\rangle\phantom{\Big|}}
		{\big\langle
		\tr \varphi^{2k+1}(x)\,\tr \overbar\varphi^{2k+1}(0)\big\rangle_0\phantom{\Big|}}
		\underset{\lambda \to \infty}{\sim}  \frac{8\pi^2}{\lambda}\, k(2k+1) + \ldots
		\label{finaloddodd}
\end{equation}
where the ellipses stand for sub-leading terms for large $\lambda$ and large $N$.
\subsection{Untwisted observables}
In Section~\ref{subsec:untwisted} we showed that, in the large-$N$ limit, the correlators of even traces have a simple expression in terms of the free energy $\cF$ given in (\ref{t2niratiofull}). 
The strong coupling behavior of the free energy 
has been recently studied in \cite{Beccaria:2021vuc}, where it has been found that 
\begin{align}
	\label{Fll}
		\cF \underset{\lambda \to \infty}{\sim} \mu \sqrt{\lambda} ~.
\end{align}
The scaling $\cF\sim \sqrt\lambda$ is strongly supported by a Pad\'e analysis of a long perturbative series for $\cF$
obtained by exploiting the matrix $\Xx$. Replacing the latter in (\ref{Fen}) by its ``leading order'' (LO) approximation (\ref{xs}),
one can analytically prove that $\mu^{\rm LO} = 1/(2\pi)$. A further discussion of the overall normalization $\mu$ will be 
presented in Section~\ref{sec:untw}.
Assuming the behavior in (\ref{Fll}), we can use the exact formula (\ref{t2niratiofull})
to express the correlators of multiple even traces at strong coupling as follows:
\begin{align}
	\label{t2nisc}
		\frac{\big\langle \tr a^{2k_1}\,\tr a^{2k_2}\cdots
		\big\rangle\phantom{\Big|}}{\big\langle \tr a^{2k_1}\,
		\tr a^{2k_2}\cdots\big\rangle_0
		\phantom{\Big|}} \underset{\lambda \to \infty}{\sim} 
		 1 - \mu\,\frac{\sum_l \,k_l(k_l+1)}{2N^2} \,\sqrt{\lambda} +\ldots~.
\end{align}

Likewise, for the correlators of the normal-ordered operators $O_{2k}(a)$ that represent the
chiral and anti-chiral operators of the gauge theory, we can use the exact formulas (\ref{O2O2exact})-(\ref{deltak}) and conclude that
\begin{equation}
\frac{\big\langle
		\tr \varphi^{2k}(x)\,\tr \overbar\varphi^{2k}(0)\big\rangle\phantom{\Big|}}
		{\big\langle
		\tr \varphi^{2k}(x)\,\tr \overbar\varphi^{2k}(0)\big\rangle_0\phantom{\Big|}}
		\underset{\lambda \to \infty}{\sim} 1 -\mu\,\frac{k(2k-1)(2k+1)}{2N^2} \,\sqrt{\lambda}
		+\ldots~.
		\label{O2kO2strong}
\end{equation}
This expression has been checked for $k=1,2,3$ and is conjectured for higher values of $k$.

\section{Numerical analysis}
\label{sec:numerical}
In this section we discuss two independent numerical methods to evaluate the 
vacuum expectation value of a given observable in the $\mathbf{E}$ theory in the 't Hooft limit at strong coupling, {\it{i.e.}} for $\lambda \gg 1$. 

The first method relies on the fact that in the weak coupling regime these vacuum expectation values admit a perturbative 
expansion in $\lambda$ that can be computed starting from their representation based on the matrix
$\Xx$ defined in (\ref{Xx}). 
These expansions may be pushed to very high order in $\lambda$ and their numerical coefficients can be evaluated with high precision. Therefore, analytic continuation beyond the convergence radius can be done by 
standard Pad\'e extrapolation \cite{Bender-Orszag,Baker}.

A different approach is based on the direct numerical integration of the matrix model provided by supersymmetric localization (see the recent analysis in \cite{Beccaria:2021ksw}). 
The vacuum expectation value of a given observable is computed by a finite $N$-dimensional integral with non-negative integrand. In the large-$N$ limit, the relevant integration domain where the integrand is sizable shrinks around the saddle point configuration, {\it{i.e.}} around a zero-measure set. This forbids direct numerical quadrature methods. Nevertheless, this is a standard problem in 
statistical mechanics and lattice field theory and its solution requires dedicated Monte Carlo (MC) methods that are well established (see for example \cite{Rothe:1992nt}).

As we will discuss in the following, the application of these two methods leads to strong coupling 
results in agreement with the analytical predictions previously discussed. 

\subsection{Methods}
\subsubsection{Pad\'e approximants for $\Delta_{1}$ and $\Delta_{2}$}

As we have seen in Section \ref{sec:leading}, it is possible to compute the quantity $\Delta_{k}$ 
defined in (\ref{fiome}) to very high orders in perturbation theory, finding
\begin{align}
    \Delta_{k} =  \sum_{n=2k+1}^{+\infty}c^{(k)}_{n} 
    \Big(\frac{\lambda}{\pi^2}\Big)^{n}~.
    \label{Deltak}
\end{align}
For example, for $\Delta_{1}$ and $\Delta_{2}$ the first few coefficients of the above expansions 
are given in \eqref{Delta1} and \eqref{Delta2}.
As it was already shown in \cite{Beccaria:2020hgy}, the radius of convergence of the series 
for $\Delta_{1}$ is located at $\lambda=\pi^2$. This is due to a branch point 
at $\lambda = -\pi^{2}$. To reveal possible higher singularities it is convenient to 
make a conformal map before computing Pad\'e approximants \cite{Costin:2020pcj,Costin:2020hwg,Costin:2019xql}. In particular, as in \cite{Beccaria:2021vuc}, we 
map the expansions (\ref{Deltak}) into the unit disk $|z|\le 1$ by the transformation
\begin{align}
\label{num1}
z = \frac{\sqrt{1+\frac{\lambda}{\pi^{2}}}-1}{\sqrt{1+\frac{\lambda}{\pi^{2}}}+1}~.
\end{align}
More precisely, we make the replacement $\frac{\lambda}{\pi^{2}} \to  \frac{4z}{(1-z)^{2}}$
in the perturbative series expansion, then we construct the Pad\'e approximants and finally we
replace $z\to \lambda$ according to (\ref{num1}).

\begin{center}
\begin{figure}
\includegraphics[width=0.45\textwidth]{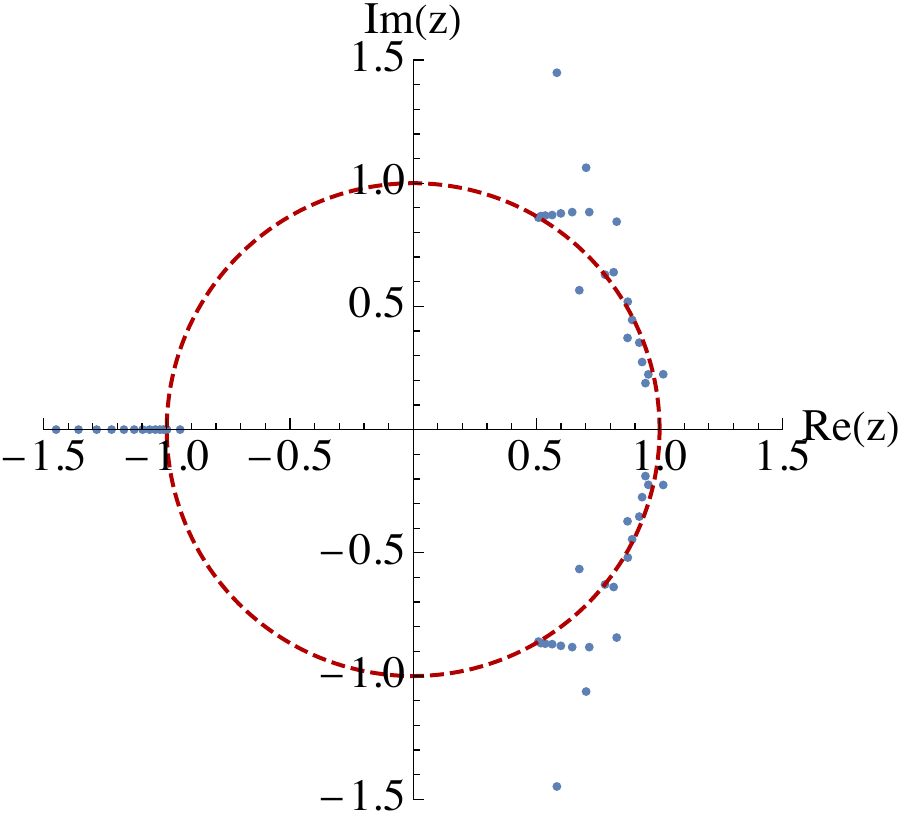}
\includegraphics[width=0.45\textwidth]{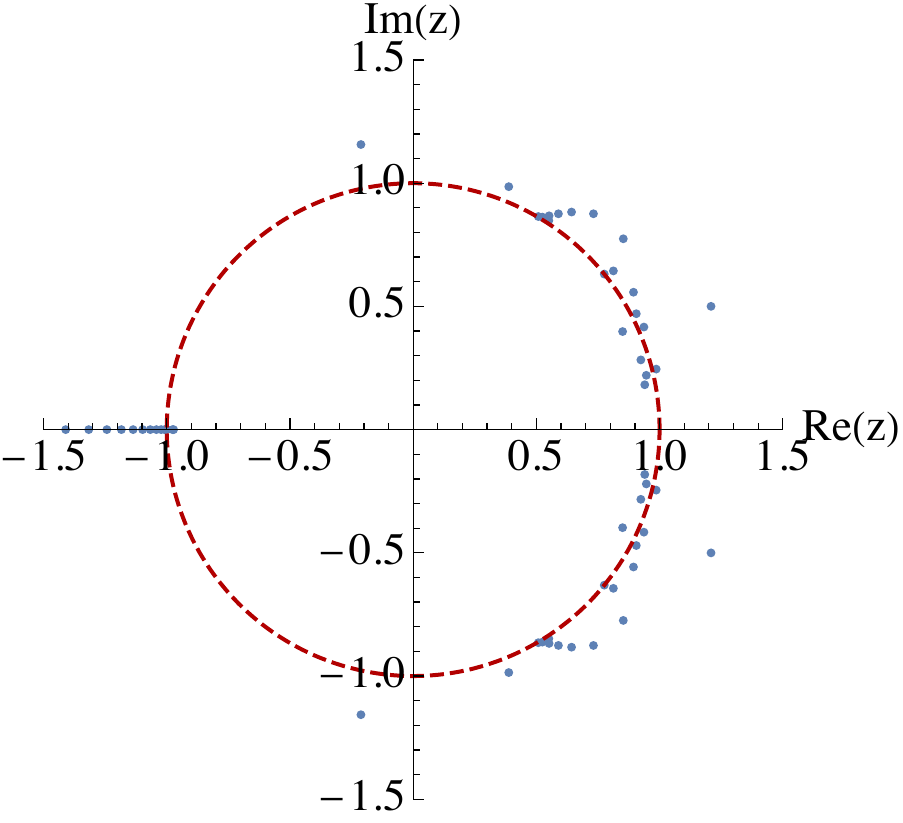}
\caption{Poles of the diagonal Pad\'e approximant of highest order for $\Delta_{1}$ (left) and $\Delta_{2}$ (right). The poles are in the $z$ variable,
after the transformation (\ref{num1}).
\label{fig:gamma35_zeroes}}
\end{figure}
\end{center}

Since we expect $\Delta_{k}$ to be asymptotically constant with $1/\lambda$ corrections, a suitable choice for the Pad\'e approximants is to take the diagonal ones. 
Hence, we shall extract information on the non-perturbative region 
$\lambda > \pi^2$, from 
\begin{align}
    P_{[M/M]}(\Delta_{k}) = \bigg[\sum_{n=2k+1}^{L}c^{(k)}_{n}\,\Big(\frac{\lambda}{\pi^2}\Big)^{n}
    \bigg]_{[M/M]}~,
\end{align}
or by improving convergence using the above conformal map.
Notice that $M$, the degree of the polynomials, must satisfy $M < L/2$. In the following we focus on the cases $k=1,2$ corresponding to $\Delta_1$ and $\Delta_2$, respectively.
In the $z$-plane, the poles of the diagonal Pad\'e approximants are shown in 
Fig.~\ref{fig:gamma35_zeroes} for $\Delta_{1}$ (left) and $\Delta_{2}$ (right). They 
tend to accumulate at $z=-1$ corresponding to the branch point
$\lambda = -\pi^{2}$.
We can observe also two symmetric sub-sequences of zeroes that tend to a point on the unit
circle corresponding to $\lambda = -4\,\pi^{2}$. The same analysis on longer perturbative expansions is expected to reveal collinear branch points at all $\lambda = -n^{2}\pi^{2}$ with 
$n=1, 2, \dots$. 

\subsubsection{Monte Carlo simulation}

There are several options for the choice of a MC algorithm suitable to our problem. 
For our purposes, and with the level of precision we will need in the analysis, 
it is possible to use one of the simplest algorithms, {\it{i.e.}} the \textit{Metropolis-Hastings} (see for instance  \cite{brooks2011handbook}). This algorithm constructs a Markov chain  
of configurations $\{Y_{j}\}_{j=0}^{\infty}$ obeying a detailed balance and each state of this chain 
corresponds to a particular configuration $Y$ of the eigenvalues $\{a_i\}$ of the 
localization matrix model. 
Moreover, each configuration $Y_j$ is weighted by a factor $\rme^{-S(Y_j)}$, where $S(Y_j)=S_{\textrm{classical}}(Y_j)+S_{\textrm{int}}(Y_j)$. 
The transition dynamics of the chain is built performing in an iterative way the two following steps:
\begin{enumerate}
    \item Starting from the element $Y_j$ we generate the next element of the chain $Y_{j+1}$. 
    This step is performed selecting in a random way one of the eigenvalues of the starting 
    configuration $Y_j$, 
    let's denote it by $\overline{a}_j$. Then we randomly change the value of $\overline{a}_j$ 
    by a step of chosen size $\epsilon$ namely
    \begin{align}
        \overline{a}_j \mapsto u\, \overline{a}_j ~,
     \end{align}
     where $u$ is a uniform random number in $[-\epsilon, \epsilon]$. 
    \item If $S(Y_{j+1}) < S(Y_j)$, the new configuration $Y_{j+1}$ is accepted. 
    While if $S(Y_{j+1}) > S(Y_j)$ the new configuration $Y_{j+1}$ is accepted only with 
    probability $\rme^{S(Y_j)-S(Y_{j+1})}$.
\end{enumerate}
In this way we obtain a large set of configurations $\{Y_j\}$  with $j=1,\cdots,n$ that are asymptotically distributed according to $\rme^{-S}$ in the limit $n\to \infty$. 
The statistical estimator of the vacuum expectation value of a generic observable $\mathcal{O}(Y_j)$ is simply the arithmetic average over the elements of the chain, namely
\begin{align}
\label{eq:vevMC}
    \big\langle \mathcal{O}(Y_j) \big\rangle = 
    \lim_{n\to \infty}\Big(\frac{1}{n}\,\sum_{j=1}^{n}\mathcal{O}(Y_j) \Big)~.
\end{align}
Evaluating (\ref{eq:vevMC}) at finite $n$ gives an estimate which has a non-negligible statistical 
error, but is exact for infinite statistics.  For finite sampling sizes, the statistical error is determined
from the variance of the estimator, corrected by the auto-correlation length of the measurements. Both these sources of errors have been measured and taken into account. As a common rule of thumb, 
we tune the size of the step $\epsilon$ in order to have an acceptance rate around $50$-$60\%$. 

\subsection{Numerical results}

\subsubsection{Twisted operators}

\begin{figure}
\center{\includegraphics[scale=0.5]{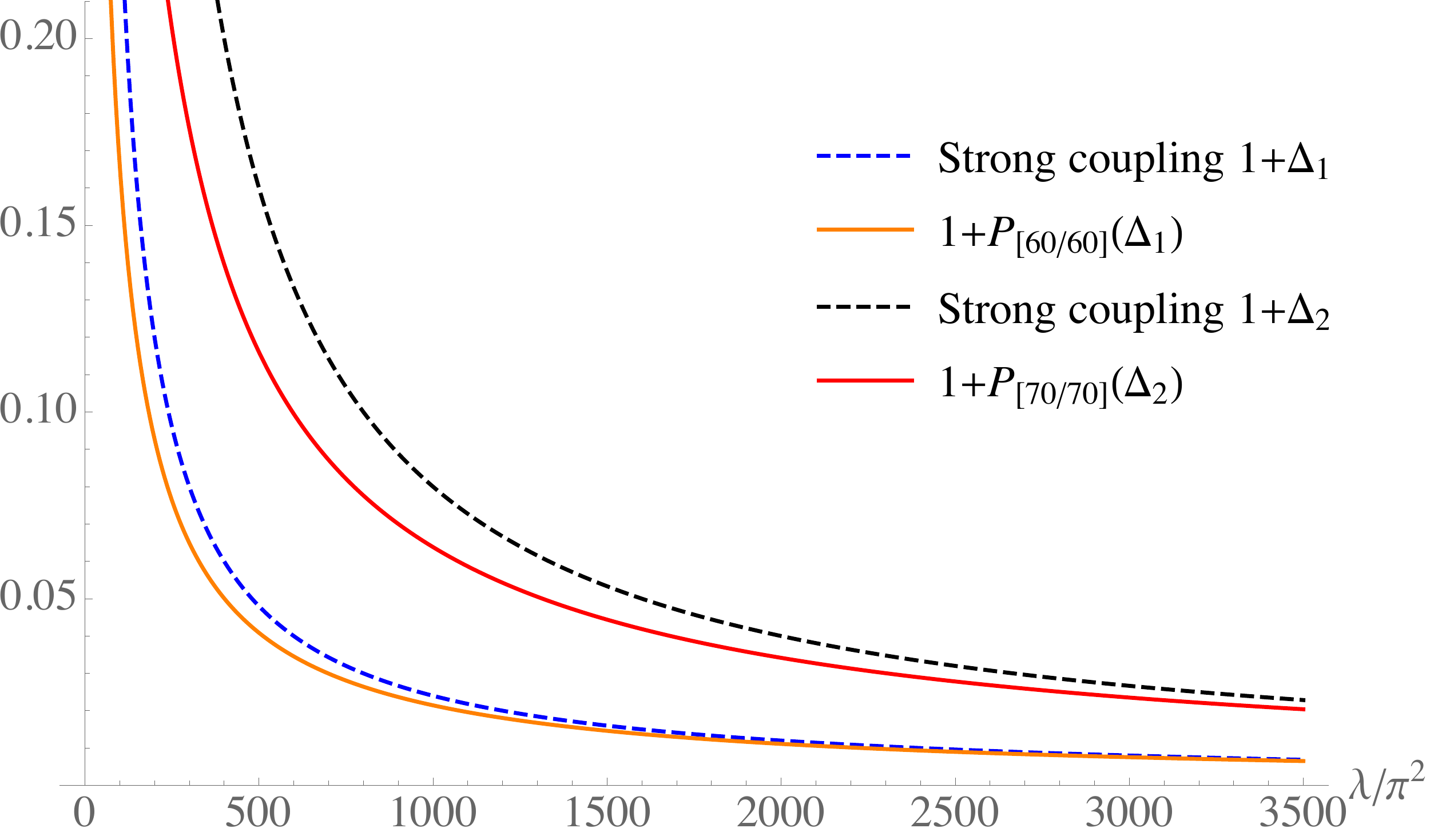}
\caption{Comparison between the Pad\'e curve $1+P_{[60/60]}(\Delta_1)$ (orange curve), the Pad\'e curve $1+P_{[70/70]}(\Delta_2)$ (red curve) and the large $\lambda$ theoretical predictions for $\gamma_3=1+\Delta_1(\lambda)$ (blue dashed curve) and $\gamma_5=1+\Delta_2(\lambda)$ (black dashed curve).\label{fig:overview}} 
    }  
\end{figure}

We first  consider the twisted observables $\Delta_{1}$ and $\Delta_{2}$ for which the analytic prediction reads (see (\ref{D3D5})):
\begin{align}
\label{eq:sc-prediction}
\gamma_3\,\equiv\,1+\Delta_{1}(\lambda) = \frac{24\,\pi^{2}}{\lambda}+\cdots~, \qquad 
\gamma_5\,\equiv\,1+\Delta_{2}(\lambda) = \frac{80\,\pi^{2}}{\lambda}+\cdots~.
\end{align}
Numerical results for these two quantities are reported in Fig. \ref{fig:overview}, where we 
compare the predictions (\ref{eq:sc-prediction}) with the diagonal Pad\'e approximants computed with $M=60$ for $\Delta_1$ and $M=70$ for $\Delta_2$. The agreement at large $\lambda$
is excellent.

\begin{center}
\begin{figure}[ht]
\includegraphics[width=0.49\textwidth]{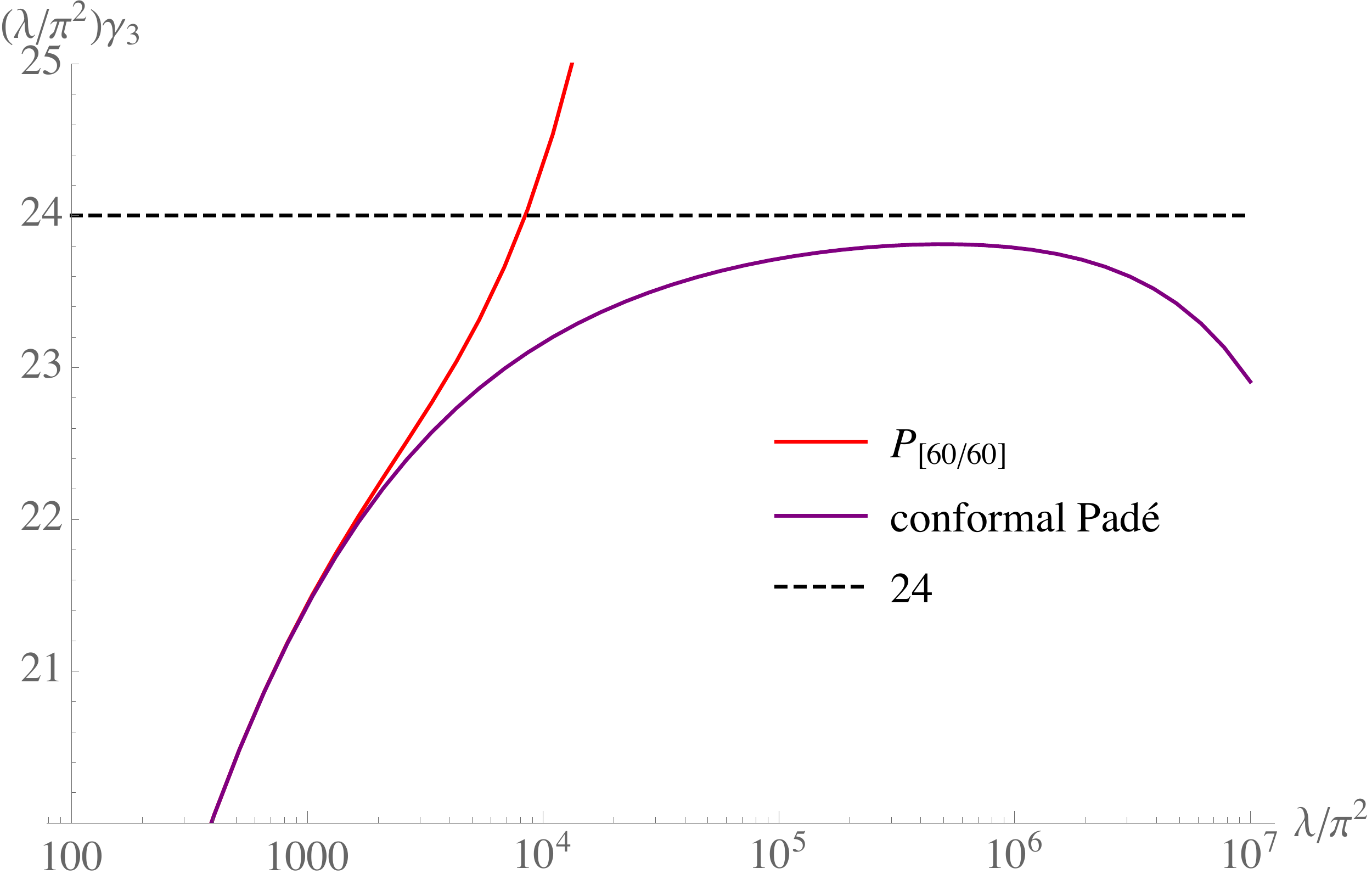}
\includegraphics[width=0.49\textwidth]{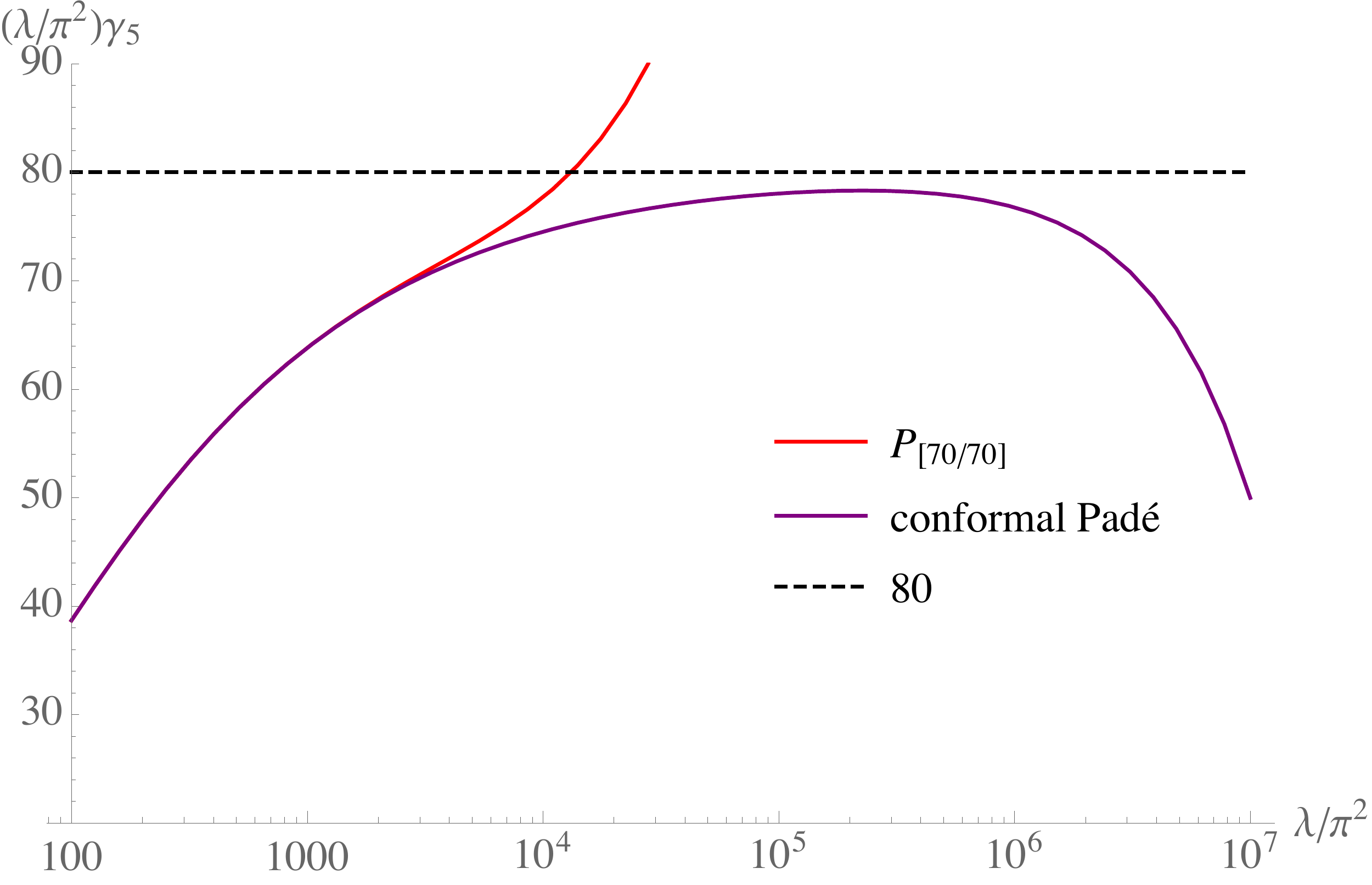}
\caption{Comparison between the simple and the conformally improved Pad\'e approximants for 
$({\lambda}/{\pi^{2}})\gamma_{3}$ (left) and $({\lambda}/{\pi^{2}})\gamma_{5}$ (right).
This quantity is expected to tend to 24 for $\gamma_{3}$ and to $80$ for $\gamma_{5}$. 
\label{fig:conformal}}   
\end{figure}
\end{center}

As we discussed, the conformal map (\ref{num1}) is effective in separating out different collinear
branch points lying on the negative real axis. Actually, it is also expected to speed up convergence with a certain fixed number of terms in the series (\ref{Deltak}). This is a general feature of a broad class of divergent series, (see for example \cite{Costin:2020hwg}),
and remains valid here where the perturbative series has a finite radius of convergence with a branch point on the boundary of the 
convergence disk. \footnote{In the divergent case, one has to apply 
the conformal map technique after Borel transformation to a convergent series.}
To appreciate the improvement associated with the conformal map (\ref{num1}), we also show in 
Fig.~\ref{fig:conformal} the products $\frac{\lambda}{\pi^2}\gamma_{3}$ and $\frac{\lambda}{\pi^2}\gamma_{5}$
using this further refinement. In the first part of these curves we see how the Pad\'e approximant tends to the right value according to (\ref{eq:sc-prediction}), but does not reach the expected plateau 
at very large $\lambda$ where it overshoots and eventually breaks down. A much better convergence is observed for the conformally improved Pad\'e approximants that come very
close to the analytical prediction before becoming unreliable. 

\begin{figure}
\center{\includegraphics[scale=0.33]{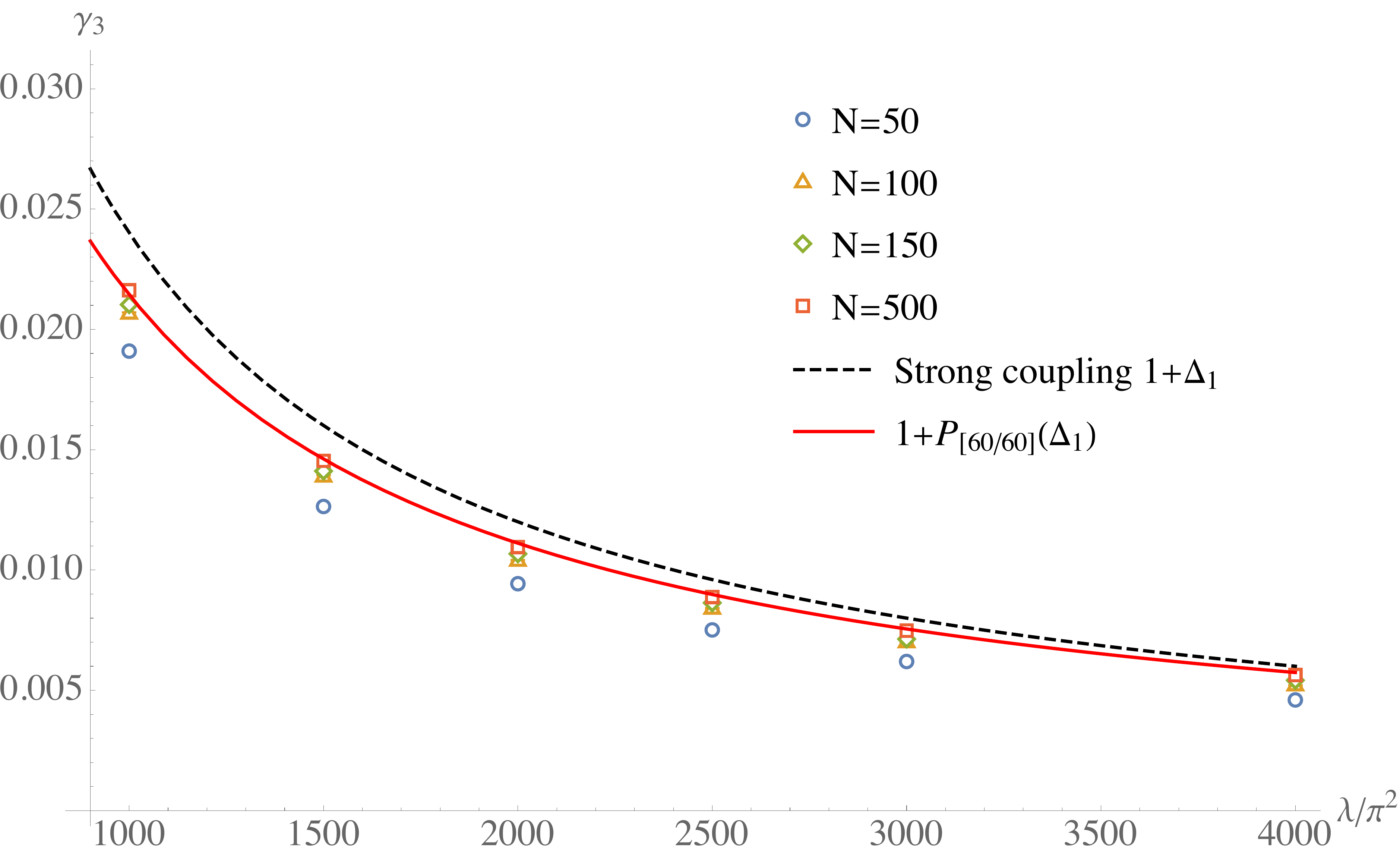}
\caption{Pad\'e curve (red curve), MC data and the large-$\lambda$ theoretical prediction for 
$\gamma_3=1+\Delta_{1}(\lambda)$ (black dashed curve) in the range 
$1000\pi^2 \leq \lambda \leq 4000\pi^2$.
As $N$ increases, the MC points systematically tend towards the Pad\'e curve.
\label{fig:gamma3_two}}}    
\end{figure}

\begin{figure}
\center{
\includegraphics[scale=0.30]{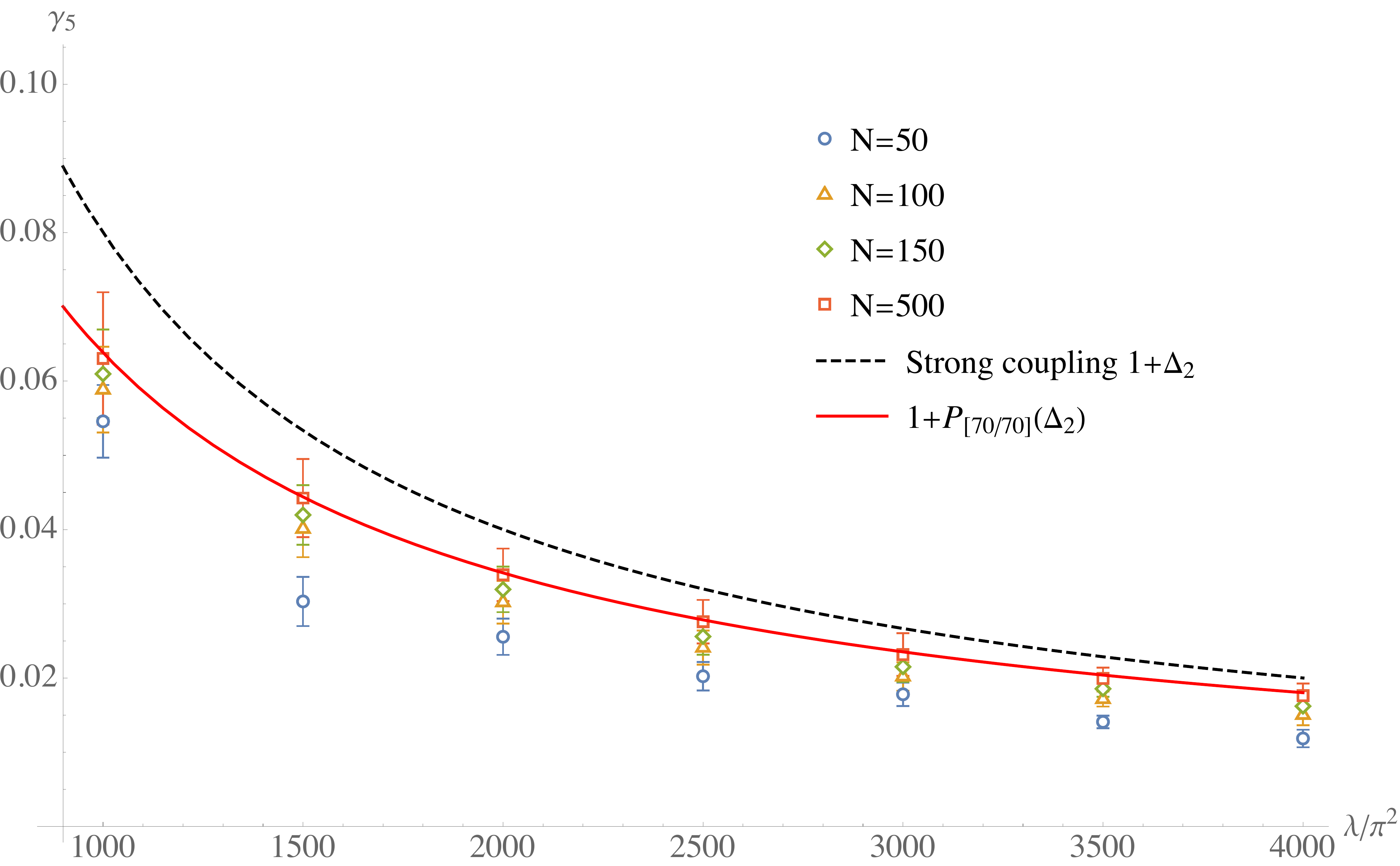}
\caption{Pad\'e curve (red curve), MC data and the large-$\lambda$ theoretical prediction for 
$\gamma_5=1+\Delta_{2}(\lambda)$ (black dashed curve) in the range $1000\pi^2 \leq \lambda \leq 4000\pi^2$. As $N$ increases, the MC points systematically tend towards the Pad\'e curve.
\label{fig:gamma5_two}}
    }
\end{figure}

The same quantities have been analyzed by MC simulations at various pairs ($N$, $\lambda$). In principle, one should extrapolate the finite $N$ results at fixed $\lambda$ in the limit $N\to \infty$. In practice, showing simultaneously our data at increasing values of $N$ clearly shows convergence. Indeed, the data points very nicely tend to the Pad\'e prediction as $N$ increases. In the MC simulations we cannot reach very large values of $\lambda$ and therefore in the comparison we use the simple Pad\'e approximants which are equivalent to the conformally improved ones in this range of coupling. This analysis is shown in 
Figs.~\ref{fig:gamma3_two} and \ref{fig:gamma5_two} for the quantities $\gamma_3=1+\Delta_{1}(\lambda)$ and $\gamma_5=1+\Delta_{2}(\lambda)$, respectively.

\subsubsection{Untwisted operators}
\label{sec:untw}

Let's now consider an application of these numerical methods to the untwisted operators. 
Although both the Pad\'e approximants and the MC simulations can be applied to a vast set of untwisted observables, for definiteness here we focus just on a particular example, namely the ratio
\begin{align}
\label{eq:h4}
\frac{\big\langle \tr a^4 \big\rangle\phantom{\Big|}}{\big\langle \tr a^4 \big\rangle_{0}\phantom{\Big|}} = 1 + \frac{q_{4}(\lambda)}{N^2} + \ldots~,
\end{align}
where, as can be seen from (\ref{t2nratiofull}), the function $q_{4}(\lambda)$ is given in terms of the free energy $\cF$ according to
\begin{equation}
q_4(\lambda)=-3\,\lambda\partial_\lambda\cF~.
\end{equation}
Therefore, in order to make an accurate prediction it is crucial to have a precise evaluation of 
$\cF$ at strong coupling. This is far from trivial as we see in the following paragraph. 

\paragraph{Strong coupling limit of $\mathcal{F}$}
A first evaluation of the free energy has been recently performed in \cite{Beccaria:2021vuc} by considering the LO expression of the matrix $\Xx$ at large $\lambda$ given in
(\ref{mtrx}). This gives the following result:
\begin{align}
\label{eq:BDT}
\mathcal{F}^{\mathrm{LO}}\underset{\lambda \to \infty}{\sim} \frac{1}{2\pi}\sqrt{\lambda} + \cdots ~.
\end{align} 
As explained in \cite{Beccaria:2021vuc}, the comparison between (\ref{eq:BDT}) and a conformally improved Pad\'e analysis is not fully satisfactory. In fact, the resummation clearly shows a 
$\sim \sqrt{\lambda}$ behavior, but the overall coefficient is sizeably smaller then the prediction 
in (\ref{eq:BDT}).
The apparent mismatch may be due to logarithmic corrections that affect convergence to the strong coupling regime but are hard to be visible in the explored range of $\lambda$ accessible by numerical simulations. Including the next-to-leading-order (NLO) term in the matrix elements of
$\Xx$ is problematic. 
Indeed, the NLO correction requires to evaluate the ill-defined quantity
\begin{align}
\label{eq:target}
\mathcal{F}^{\mathrm{NLO}}= \frac{1}{2}\log \det \Big(\frac{2}{3}
+\frac{\lambda}{2\pi^2}\Sx \Big) ~, 
\end{align}
where $\Sx$ is the three-diagonal matrix given in \eqref{s}.
One possibility is to evaluate \eqref{eq:target} numerically by the empirical procedure described in 
Section~6.2. of \cite{Beccaria:2021vuc}. In practice this means defining
\begin{align}
\mathcal{F}_k^{\mathrm{NLO}}
= \frac{1}{2}\,\log\det \Big(\frac{2}{3}+\frac{\lambda}{2\pi^2}\,\Sx_{(k)}\Big)
\end{align}
where, in the notation introduced after \eqref{gammaome0}, $\Sx_{(k)}$ denotes
the upper left $k \times  k$ block of $\Sx$, and then extrapolating the result for $k\to\infty$.
In this limit this provides an estimate for $\mathcal{F}^{\mathrm{NLO}}$. To obtain this
extrapolation, we first fix $k$ and look for the value of $\lambda$ that maximizes the ratio ${\mathcal{F}_k^{\mathrm{NLO}}}/{\sqrt{\lambda}}$. We denote this maximum as
\begin{align}
\mu_k^{\mathrm{NLO}}= \max _{\lambda} 
\Big(\frac{\mathcal{F}_k^{\mathrm{NLO}}}{\sqrt{\lambda}}\Big)~.
\end{align}
Then, by successively increasing $k$, we generate a set of values 
of $\mu_k^{\mathrm{NLO}}$, which we fit as follows
\begin{align}
\mu_k^{\mathrm{NLO}} = 0.113-\frac{0.164}{\sqrt{k}} - \frac{0.0677}{k}~.
\end{align}
The value 
\begin{align}
\label{eq:NLO}
\mu^{\mathrm{NLO}}= 0.113~,
\end{align}
provides an empirical estimate of the coefficient in front of $\sqrt{\lambda}$ at the NLO. This result 
is substantially smaller than the LO value ${1}/{2\pi}$ given in (\ref{eq:BDT}) by a factor of about $\sqrt 2$. 

A further refinement consists in avoiding the use of the asymptotic expansion of the 
matrix elements $\Xx_{k\ell}$ and in using instead their exact expression 
as a definite integral given in\eqref{Xx}. This is non-trivial because of the highly oscillatory behavior of the integrand associated with the Bessel functions. 
We can deal with it by splitting the integration region in sub-intervals between two zeroes of $J_{2k+1}J_{2\ell+1}$. This gives an alternating series for which it is
possible to estimate the convergence error. Repeating the analysis in this way, we have obtained
\begin{align}
\label{eq:full}
\mu^{\mathrm{full}}=  0.130~,
\end{align}
where ``full'' emphasizes that we used the exact (numerical) values of 
$\Xx_{k\ell}(\lambda)$. The extrapolation in (\ref{eq:full}) is done with smaller values of $k$ than in (\ref{eq:NLO}), but the shown three digits appear to be stable.

\begin{figure}
\center{\includegraphics[scale=0.45]{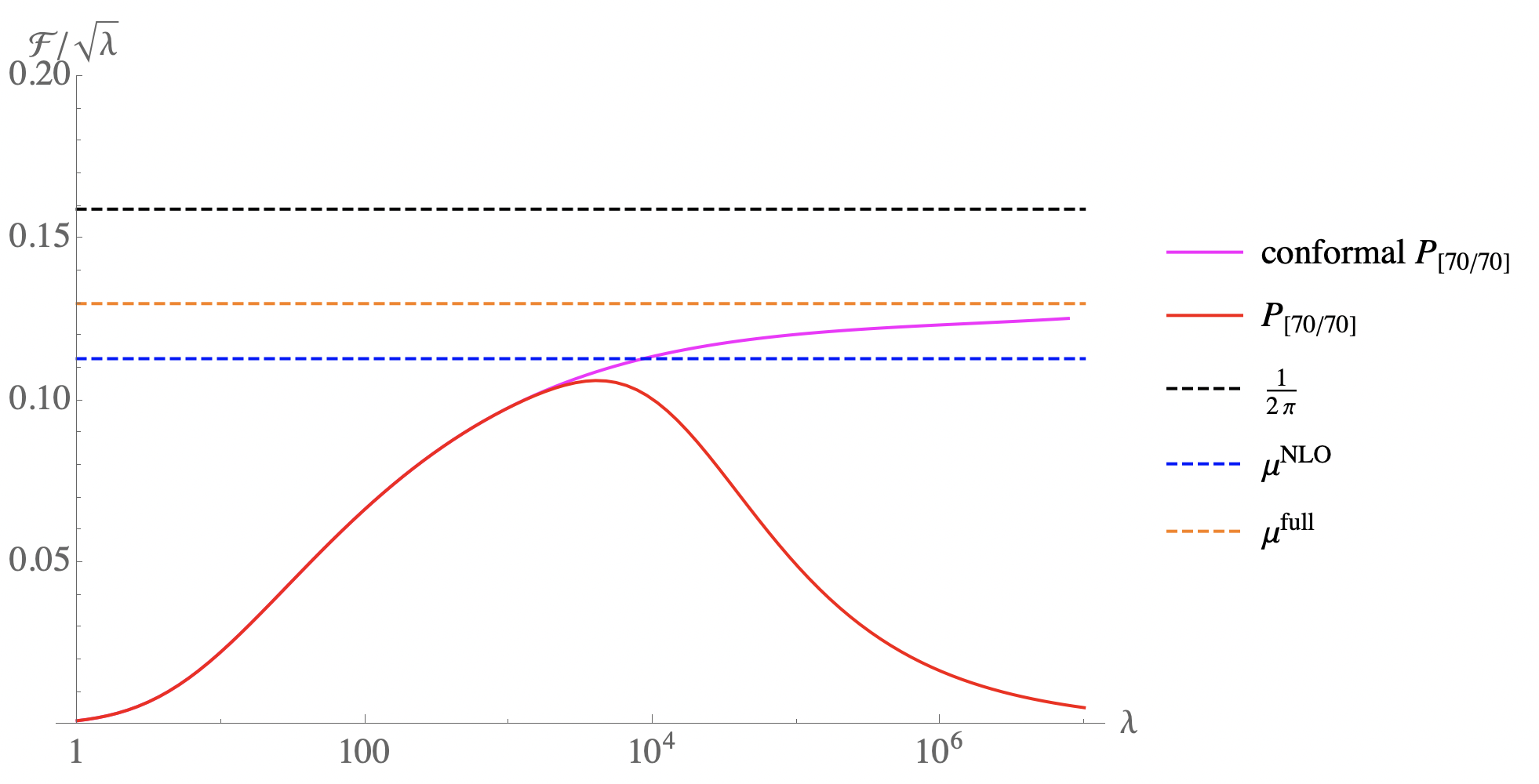}
\caption{
Comparison between various estimations of $\mathcal{F}/\sqrt\lambda$. The highest horizontal line is the expected asymptotic constant
using the LO analytical prediction. The other horizontal lines are the prediction from the truncated matrix analysis using the NLO expansion of the $M$ matrix elements or  their ``full'' integral 
representation. Also shown are the simple and conformally improved Pad\'e resummations taken from \cite{Beccaria:2021vuc}.
 Notice that a residual steady increase of the purple curve at quite large values of 
 $\lambda$ is clearly visible in the adopted logarithmic scale.
 \label{fig:comparison}.} 
 }  
\end{figure}

In Fig.~\ref{fig:comparison} we show the Pad\'e 
resummation of $\mathcal{F}/\sqrt{\lambda}$ and its conformal improvement, taken from \cite{Beccaria:2021vuc}, together with the three estimates for the asymptotic value of this ratio, {\it{i.e.}} the 
LO prediction $1/2\pi$ from (\ref{eq:BDT}), the NLO value in (\ref{eq:NLO}), and the ``full'' extrapolation in (\ref{eq:full}). In the adopted logarithmic scale, we can reveal possible logarithmic
contributions that may substantially slow down the convergence to the plateau. The conformally improved purple curve overshoots the NLO prediction, while it remains below $\mu^{\mathrm{full}}$
almost reaching it at $\lambda \sim 10^{7}$. In the range $\lambda < 10^{3}$ the curve is clearly not flat, but totally under control, given the agreement between the 
simple and the conformally improved Pad\'e approximants.

\begin{figure}
    \centering
    \includegraphics[scale=0.5]{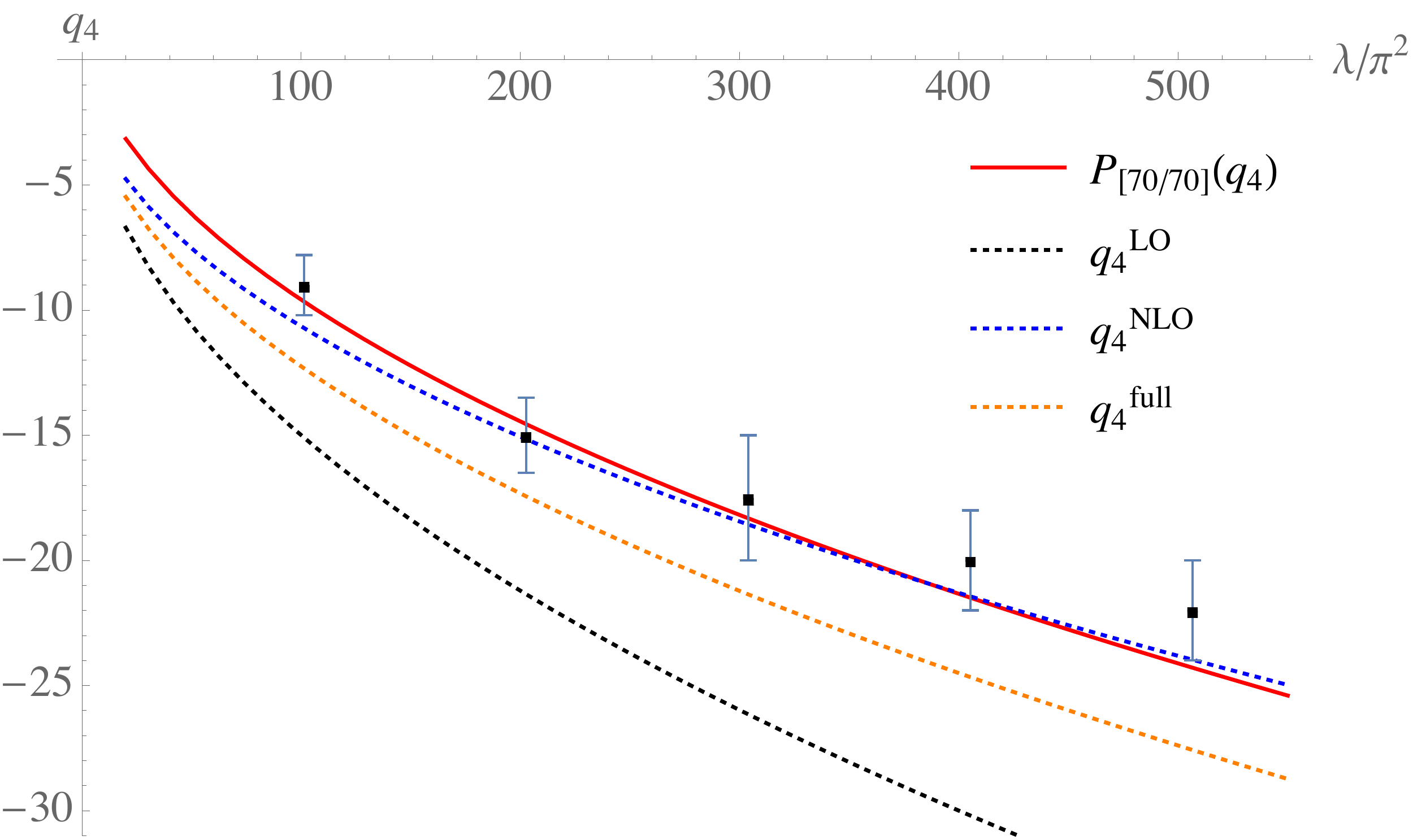}
    \caption{Comparison between the diagonal Pad\'e approximant $P_{[70/70]}(q_4)$ (red curve), the MC points and the functions $q^{\mathrm{LO}}_4(\lambda)$, $q^{\mathrm{NLO}}_4(\lambda)$, $q^{\mathrm{full}}_4(\lambda)$ defined in \eqref{eq:tnew} and \eqref{eq:textra}. We observe the agreement between the Pad\'e approximant $P_{[70/70]}(q_4)$  and the MC points (within error bars). Since, as shown in Fig. \ref{fig:comparison}, the asymptotic regime becomes manifest only for larger values, i.e. $\lambda \sim 10^6$, we do not expect the expressions $q^{\mathrm{LO}}_4(\lambda)$, $q^{\mathrm{NLO}}_4(\lambda)$, $q^{\mathrm{full}}_4(\lambda)$ to be a good approximation in the range $0 \leq \lambda/\pi^2 \leq 550$ analyzed here.}
    \label{fig:all} 
\end{figure}

\paragraph{Analysis of $q_{4}(\lambda)$}
Let's now return to  function $q_{4}(\lambda)$ appearing in \eqref{eq:h4}. 
Using \eqref{eq:BDT} for the strong-coupling behavior of the free energy, 
we obtain
\begin{align}
\label{eq:tnew}
q_4^{\mathrm{LO}}(\lambda)\underset{\lambda \to \infty}{\sim} -\frac{3}{2}\sqrt{\frac{\lambda}{\pi^2}}+\ldots~.
\end{align}
If instead we use the estimates \eqref{eq:NLO} and \eqref{eq:full}, we 
have
\begin{equation}
\label{eq:textra}
\begin{aligned}
q^{\mathrm{NLO}}_4(\lambda) &\underset{\lambda \to \infty}{\sim}
 2\pi\mu^{\mathrm{NLO}} \times q_4^{\mathrm{LO}}(\lambda)+\ldots~, \\[1mm]
q^{\mathrm{full}}_{4}(\lambda) &\underset{\lambda \to \infty}{\sim}
 2\pi\mu^{\mathrm{full}} \times q_4^{\mathrm{LO}}(\lambda)+\ldots~.
\end{aligned}
\end{equation}
We have compared these predictions with two independent numerical evaluations, namely a diagonal Pad\'e approximant\,%
\footnote{In the considered range of $\lambda$, the simple Pad\'e approximant
and its conformally improved version are substantially the same. Thus for simplicity we
make the comparison with the Pad\'e approximant.} 
and a MC simulation. Our results are collected
in Fig.~\ref{fig:all} where we observe equality  between the Pad\'e approximant and the MC data
within error bars. Given the considered large values of $\lambda$, this is an important independent 
check  of the proposed resummation based on the Pad\'e representation.
However, as it is clear from Fig.~\ref{fig:comparison}, we do not expect  the
expressions (\ref{eq:tnew}) and (\ref{eq:textra}) to be a good approximation in the range explored in Fig.~\ref{fig:all}.
Indeed, the asymptotic regime becomes visible at  much larger values $\lambda\sim 10^{6}$,
presumably due to slow logarithmic corrections. Nevertheless, it may be interesting to look at 
the behavior of the three analytical expressions in (\ref{eq:tnew}) and(\ref{eq:textra}).
We observe that the LO prediction \eqref{eq:tnew} is systematically much lower than the MC data and the Pad\'e approximant. The NLO curve starts to overshoot 
the Pad\'e approximant at $\lambda \sim 350 \pi^2$, while the ``full'' curve closely follows the Pad\'e resummation keeping underneath it, but with similar slope.

We remark that the untwisted MC measurements that we could achieve still have sizeable relative errors around $10\%$. Besides, the data at higher values of $\lambda$
are less reliable due to systematic errors in the large-$N$ extrapolation needed to extract 
$q_{4}(\lambda)$ from the slope in (\ref{eq:h4}).
This fact is an indication that the untwisted observables, as the one we have considered, are much more difficult to be evaluated via a MC simulation than the twisted ones. 
In order to reduce the relative errors and increase the precision, a more intense computational effort would be necessary. However this is far beyond the scope of the present work. 
As a final comment, we stress the great flexibility of the MC approach. In this framework, in fact,
more complicated observables may be studied at quite large values of gauge coupling with minor effort, even without resorting to the special role played by the $\Xx$ matrix in the considered model.  

\vskip 1cm
\noindent {\large {\bf Acknowledgments}}
\vskip 0.2cm
We would like to thank Lorenzo Bianchi, Gerald V. Dunne, Francesco Galvagno and Igor Pesando for helpful discussions.
This research is partially supported by the INFN projects ST\&FI
``String Theory \& Fundamental Interactions'' and GSS ``Gauge, Strings and Supergravity''. The work of A.P. is supported by INFN with a``Borsa di studio post-doctoral per fisici teorici".
\vskip 1cm
\begin{appendix}

\section{Details on the derivation of the strong-coupling results}
\label{app:details}
In Section~\ref{sec:large} we have written in (\ref{xs}) the leading behavior for large $\lambda$ of the infinite matrix $\Xx$ in terms of the infinite matrix $\Sx$ defined in (\ref{s}). At strong coupling, the functions $\Delta_k(\lambda)$ that represent the deviation from the $\cN=4$ result of the correlator of two twisted observables can then be expressed in terms of determinants of the sub-matrices of $\Sx$, as shown in (\ref{detDdetS}). Moreover, the expectation values of odd multi-traces are expressed 
in terms of the elements of the matrix $\Dx$ and thus, using (\ref{Dll}), in terms of the elements of the inverse of $\Sx$. In this appendix we give some details on how to treat these infinite matrices
and compute the invariants connected with $\mathsf{S}$.

We first perform a similarity transformation and write
\begin{equation}
	\label{sdy}
		\mathsf{S} =\mathsf{T} \mathsf{Y} \mathsf{T}^{-1}~,
\end{equation}
where $\mathsf{T}$ is the diagonal matrix 
\begin{equation}
	\label{dij}
		\mathsf{T}_{k\ell} = (-1)^k \, \sqrt{2k+1} \, \delta_{k,\ell}
\end{equation}
and $\Yx$ is not symmetric but has rational entries given by 
\begin{align}
	\label{y}
		\mathsf{Y}_{k\ell}	=
		\frac{1}{4} 
		\Big(\frac{\delta_{k-1,\ell}}{k (2k+1)} + \frac{\delta_{k,\ell}}{k(k+1)} +
		\frac{\delta_{k+1,\ell}}{(k+1)(2k+1)}
		\Big)~. 
\end{align}
Of course we have
\begin{align}
	\label{StoYvaria}
		\Sx^{-1} = \mathsf{T} \mathsf{Y}^{-1} \mathsf{T}^{-1}~,~~~
		\det \mathsf{S}= \det \mathsf{Y}~,~~~
		\tr \mathsf{S}^n= \tr \mathsf{Y}^n~.
\end{align}
Note that such relations are valid also for the infinite sub-matrices in which the first $k-1$ rows and columns are removed, so that in particular
\begin{align}
	\label{detSkYk}
		\det \Sx_{[k]} = \det \Yx_{[k]}~.
\end{align} 
The matrix $\mathsf{Y}$, in turn, can be written as 
\begin{equation}
\mathsf{Y} =\mathsf{N}\,\mathsf{\Lambda}~,
\label{yl}
\end{equation}
where $\mathsf{N}$ is the diagonal matrix with entries
\begin{align}
	\label{Nis}
		\mathsf{N}_{k,\ell}= \frac{\delta_{k,\ell}}{2k(2k+1)(2k+2)}~,
\end{align}
while $\mathsf{\Lambda}$ is a three-diagonal integer matrix with entries
\begin{align}
\label{Lambdais}
\mathsf{\Lambda}_{k,\ell}= (k+1)\,\delta_{k-1,\ell} + (2k+1)\,\delta_{k,\ell} +
k\,\delta_{k+1,\ell} ~.
\end{align}
Therefore, for every $k\geq 1$ we have
\begin{align}
	\label{detYkNk}
		 \det \Yx_{[k]} = \det \Nx_{[k]} \det \Lax_{[k]}~.
\end{align}
These infinite determinants \emph{per se} are ill-defined and should be regularized. However the ratios 
\begin{equation}
\frac{\det\Yx_{[k+1]}}{\det\Yx_{[k]}}~,
\end{equation}
which are the quantities that determine the observables $\Delta_k(\lambda)$ according to (\ref{detDdetS}), are not only well-defined but have a simple expression that we shall now derive. 

The determinant of $\Nx_{[k]}$ is formally given by
\begin{align}
	\label{detNk}
		\det \Nx_{[k]} = \prod_{\ell=k}^{\infty} \left(\frac{1}{2\ell(2\ell+1)(2\ell+2)}\right)~.
\end{align}
To compute $\det\Lax_{[k]}$ let us start from the case $k=1$, {\emph{i.e.}} from $\det\Lax$. 
The three-diagonal matrix $\Lax$ is
\begin{equation}
\mathsf{\Lambda}=
\begin{pmatrix}
3&1&0&0&0&0& \\
3&5&2&0&0&0& \\
0&4&7&3&0&0& \cdots \\
0&0&5&9&4&0& \\
0&0&0&6&11&5& \\
0&0&0&0&7&13& \ddots\\
&& \vdots &&&\ddots & \ddots
\end{pmatrix}
\label{matrla}
\end{equation}
The determinant of $\Lax$ is unchanged if we substitute $\Lax$ with a matrix $\Lax^\prime$ whose rows are linear combinations of the rows of $\Lax$. If we choose each row of $\Lax^\prime$ to be the alternating sum of all preceding rows of $\Lax$, namely if we set
\begin{align}
	\label{Laxpis}
		\Lax^\prime_{k,\ell} = \Lax_{k,\ell} - \Lax_{k-1,\ell} + \Lax_{k-2,\ell} 
		+ \ldots - (-1)^k \Lax_{1,\ell}~, 
\end{align}
then $\Lax^\prime$ is upper-triangular with elements
\begin{align}
	\label{Laxpel}
		\Lax^\prime_{k,\ell} = (k+2) \delta_{k,\ell} + k\, \delta_{k+1,\ell}~.
\end{align}
Explicitly, we have
\begin{equation}
\Lax^\prime=
\begin{pmatrix}
3&1&0&0&0&0& \\
0&4&2&0&0&0& \\
0&0&5&3&0&0& \cdots \\
0&0&0&6&4&0& \\
0&0&0&0&7&5& \\
0&0&0&0&0&8& \ddots\\
&& \vdots &&&\ddots & \ddots
\end{pmatrix}~.
\label{matrlap}
\end{equation}
In this way we get
\begin{equation}
\det \mathsf{\Lambda} = \det \mathsf{\Lambda'} = \prod_{\ell=1}^{\infty} (\ell+2) 
\label{detl}
\end{equation}
Using this result together with (\ref{detNk}) for $k=1$, from (\ref{detYkNk}) we get
\begin{equation}
\det \mathsf{Y} = \det \mathsf{Y}_{[1]} = \frac{1}{2}
\prod_{\ell=1}^{\infty} \Big(\frac{1}{2\,(2\ell+1)(2\ell+2)}\Big)~.
\label{detty}
\end{equation}
With the same technique we can also compute the determinants of the matrices 
\begin{equation}
\Lax_{[k]}=
\begin{pmatrix}
2k+1&k&0&0&0& \\
k+2&2k+3&k+1&0&0& \\
0&k+3&2k+5&k+2&0& \cdots \\
0&0&k+4&2k+7&k+3& \\
0&0&0&k+5&2k+9& \ddots\\
&& \vdots &&\ddots & \ddots
\end{pmatrix}
\label{matrlak}
\end{equation}
Again, the determinant of $\Lax_{[k]}$ is equal to the determinant of the matrix $\Lax_{[k]}^\prime$ obtained from $\Lax_{[k]}$ analogously to $\Lax^\prime$ by summing with alternating signs the
rows, which is 
\begin{equation}
\Lax_{[k]}^\prime=
\begin{pmatrix}
2k+1&k&0&0&0& \\
-k+1&k+3&k+1&0&0& \\
k-1&0&k+4&k+2&0& \cdots \\
-k+1&0&0 &k+5&k+3& \\
k-1&0&0&0&k+6& \ddots\\
&& \vdots &&\ddots & \ddots
\end{pmatrix}
\label{matrlakp}~.
\end{equation}
This matrix is not upper-triangular, but its determinant can be easily 
computed by expanding it with respect to the first column. In this way we get
\begin{align}
\det \mathsf{\Lambda}^\prime_{[k]}=& \,(2k+1) \Big(\prod_{\ell=k+3}^{\infty}\ell \Big)
+(k-1) \Big(k \prod_{\ell=k+4}^{\infty}\ell\Big)+
(k-1) \Big(k (k+1) \prod_{\ell=k+5}^{\infty}\ell\Big)+ \cdots \notag \\
=& \bigg[ (2k+1) + (k-1) \sum_{m=1}^{\infty} \frac{(k)_m}{(k+3)_m} \bigg]\,
\Big(\prod_{\ell=k+3}^{\infty}\ell\Big)~.
\end{align}
The series in the square brackets is recognized to yield an hypergeometric function, so that
\begin{align}
\det \mathsf{\Lambda}^\prime_{[k]}&
 = \Big[ k+2 + (k-1) F(k,1,k+3,1) \Big]\,\Big(
\prod_{\ell=k+3}^{\infty}\ell \Big)
=  \frac{1}{2} \prod_{\ell=k+1}^{\infty}\ell =\det \mathsf{\Lambda}_{[k]}~.
\label{detlapn}
\end{align}
Multiplying this result with (\ref{detNk}) and using (\ref{detYkNk}), we get
\begin{equation}
\det \mathsf{Y}_{[k]} =\frac{1}{2k} \,\prod_{\ell=k}^{\infty} \Big(\frac{1}{2\,(2\ell+1)(2\ell+2)}\Big)~,
\label{detynr}
\end{equation}
which generalizes (\ref{detty}).
Finally we obtain
\begin{equation}
\big(\mathsf{S}^{-1}_{[k]}\big)_{1,1} =
\frac{\det \mathsf{S}_{[k+1]}}{\det \mathsf{S}_{[k]}} 
= \frac{\det \mathsf{Y}_{[k+1]}}{\det \mathsf{Y}_{[k]}} 
= 4k \,(2k+1)~
\label{yn11}
\end{equation}
where the second equality stems from (\ref{detSkYk}). This result has been used in Eq.s (\ref{detDdetS}) and (\ref{resDeltak}) of the main text.

In the same way we can compute the determinants of the sub-matrices 
$\mathsf{Y}_{(k)}$ corresponding to the upper $(k \times k)$ block of $\mathsf{Y}$.
Indeed, defining $\mathsf{N}_{(k)}$ and $\mathsf{\Lambda}_{(k)}$ in the same way, we have
\begin{equation}
\mathsf{Y}_{(k)} =\mathsf{N}_{(k)}\,\mathsf{\Lambda}_{(k)}~,
\label{ylnfinito}
\end{equation}
and
\begin{equation}
\det \mathsf{Y}_{(k)} =\det \mathsf{N}_{(k)}\,\det \mathsf{\Lambda}_{(k)} =
\frac{1}{2} (k+1)(k+2) \prod_{\ell=1}^{k} \Big(\frac{1}{(2\ell+1)(2\ell+2)}\Big)~.
\label{detynfinito}
\end{equation}

More generally, these techniques can be used to compute any element of the infinite matrix  
$\mathsf{Y}^{-1}$.
In fact,
\begin{equation}
\mathsf{Y}^{-1}_{k,\ell} =\frac{\widehat{\mathsf{Y}}_{\ell,k}}{\det \mathsf{Y}} 
\label{ymuno}
\end{equation}
where the $\widehat{\mathsf{Y}}$ is the cofactor matrix whose entries are
\begin{equation}
\widehat{\mathsf{Y}}_{n,m} = (-1)^{n+m} \det \mathsf{Y}_{[n,m]} 
\end{equation}
with $\Yx_{[n,m]}$ being the sub-matrix obtained from $\Yx$ by deleting its $n^{\mathrm{th}}$
row and its $m^{\mathrm{th}}$ column. Since $\Yx=\Nx\,\Lax$ with $\Nx$ given
in (\ref{Nis}), we have
\begin{equation}
\widehat {\mathsf{Y}}_{n,m} = (-1)^{n+m}\,
\prod_{j \neq n}^{\infty} \,\Big(\frac{1}{2j\,(2j+1)(2j+2)}\Big)\, \det \mathsf{\Lambda}_{[n,m]}~,
\label{yminor}
\end{equation}
where the matrix $\mathsf{\Lambda}_{[n,m]}$ is defined similarly to $\mathsf{Y}_{[n,m]}$.

We observe that when $n \leq m$, the matrix $\mathsf{\Lambda}_{[n,m]} $ is upper triangular
in block form and can be written as
\begin{equation}
\mathsf{\Lambda}_{[n,m]} =
\begin{pmatrix}
\mathsf{\Lambda}_{(n-1)}& \star & 0\\
&\mathsf{U}&\star\\
&& \mathsf{\Lambda}_{[m+1]}
\end{pmatrix}~,
\label{lanmlul}
\end{equation}
where $\mathsf{U}$ is the upper-triangular matrix of dimension $(m-n)$ given by
\begin{equation}
\mathsf{U} =
\begin{pmatrix}
n+2&*&*&*&\\
0&n+3&*&*&\cdots\\
0&0&n+4&*& \\
&\vdots& &\ddots&\vdots \\
0&0&0&\cdots&m+1 
\end{pmatrix}~,
\label{Unm}
\end{equation}
whose determinant is 
\begin{equation}
\det \mathsf{U}= \prod_{\ell=n+2}^{m+1} \ell~.
\label{detUnm}
\end{equation}
On the other hand, when $n \geq m$ the matrix $\mathsf{\Lambda}_{[n,m]}$ is lower triangular
in block form and reads
\begin{equation}
\mathsf{\Lambda}_{[n,m]} =
\begin{pmatrix}
\mathsf{\Lambda}_{(m-1)}&&\\
\star &\mathsf{L}&\\
0&\star& \mathsf{\Lambda}_{[n+1]}
\end{pmatrix}~,
\label{lanmlll}
\end{equation}
where $\mathsf{L}$ is a lower triangular matrix of dimension $(n-m)$ given by
\begin{equation}
\mathsf{L}=
\begin{pmatrix}
m&0&0&&0\\
*&m+1&0&\cdots&0\\
*&*&m+2&&0 \\
&\vdots& &\ddots&\vdots \\
*&*&*&\cdots&n-1 
\end{pmatrix}~,
\label{Lnm}
\end{equation}
whose determinant is
\begin{equation}
\det \mathsf{L}= \prod_{\ell=m}^{n-1} \ell ~.
\label{detLnm}
\end{equation}
Using these results in (\ref{yminor}), we can compute all the minors of $\mathsf{Y}$, finding:
\begin{align}
\widehat {\mathsf{Y}}_{n,m} & = (-1)^{n+m} \prod_{\ell \neq n} \frac{1}{2\ell\,(2\ell+1)(2\ell+2)}
\,\det \mathsf{\Lambda}_{(n-1)}\, \det \mathsf{U}\, \det \mathsf{\Lambda}_{[m+1]}
\notag \\
& = (-1)^{n+m}\,n (n+1)(2n+1) \,\prod_{\ell=1}^{\infty} 
\frac{1}{2\,(2\ell+1)(2\ell+2)}
\label{yminor2}
\end{align}
for $n \leq m$, and
\begin{align}
\widehat {\mathsf{Y}}_{n,m} & =(-1)^{n+m} \prod_{\ell \neq n} \frac{1}{2\ell\,(2\ell+1)(2\ell+2)} 
\,\det \mathsf{\Lambda}_{(m-1)}\, \det \mathsf{L}\, \det \mathsf{\Lambda}_{[n+1]}
\notag \\
& = (-1)^{n+m}\,m(m+1)(2n+1)\, 
\prod_{\ell=1}^{\infty}  \frac{1}{2\,(2\ell+1)(2\ell+2)}
\label{yminor3}
\end{align}
for $n \geq m$.

Inserting these expressions in (\ref{ymuno}), we finally obtain
\begin{equation}
\mathsf{Y}^{-1}_{k,\ell} = 
\begin{cases}
(-1)^{k+\ell}\, 2k\,(k+1) (2\ell+1)& \quad\mbox{for}~k \leq \ell~,\\[2mm]
(-1)^{k+\ell}\, 2\ell\,(\ell+1) (2\ell+1)& \quad\mbox{for}~k \geq \ell~.
\end{cases}
\label{ymuno1app}
\end{equation}
Then, from the first equation in (\ref{StoYvaria}) we have
\begin{equation}
\mathsf{S}^{-1}_{k,\ell}=(-1)^{k+\ell}\, \sqrt{\frac{2k+1}{2\ell+1}}\,\mathsf{Y}^{-1}_{k,\ell} = 2
\sqrt{(2k+1)(2\ell+1)}
\begin{cases}
k\,(k+1) & \quad\mbox{for}~k \leq \ell~,\\[2mm]
\ell\,(\ell+1) & \quad\mbox{for}~k \geq \ell~.
\end{cases}
\label{ymuno2app}
\end{equation}
which has been used in Eq. (\ref{ymuno1}) of the main text.

\end{appendix}

%
%

\providecommand{\href}[2]{#2}\begingroup\raggedright\endgroup

\end{document}